\documentclass[onecolumn,amsmath,amssymb,11pt,superscriptaddress,nofootinbib]{revtex4-2}
\usepackage[T1]{fontenc}
\usepackage{lmodern}
\usepackage[british]{babel}
\makeatletter\AtBeginDocument{\let\@elt\relax}\makeatother
\usepackage{amsmath}
\usepackage{amssymb}
\usepackage{physics}
\usepackage{amsthm}
\usepackage[]{graphicx}
\usepackage[justification=centering]{caption}
\usepackage{subcaption}
\usepackage{tensor}
\usepackage[dvipsnames]{xcolor}
\usepackage{cancel}
\usepackage{setspace}
\usepackage{fancyhdr}
\usepackage{appendix}
\usepackage{url}
\usepackage{natbib}
\usepackage{hyperref}
\hypersetup{
	colorlinks = true,
	linkbordercolor = {white},
	linkcolor=black,
	citecolor=blue,
	urlcolor=black}
\usepackage[a4paper,margin=2.5cm]{geometry}

\begin{document}
	
	\title{Birth of Inflationary Universes via Wineglass Wormholes and their No-Boundary Relatives}
	
	\author{George Lavrelashvili}
	\email[]{george.lavrelashvili@tsu.ge}
	\affiliation{Department of Theoretical Physics, A.Razmadze Mathematical Institute \\
		at I.Javakhishvili Tbilisi State University, GE-0193 Tbilisi, Georgia}
	\author{Jean-Luc Lehners}
	\email[]{jlehners@aei.mpg.de}
	\affiliation{Max Planck Institute for Gravitational Physics \\ (Albert Einstein Institute), 14476 Potsdam, Germany}
	
\begin{abstract}
\vspace{0.5cm}
We study Euclidean wineglass wormholes, which mediate the nucleation of inflationary spacetimes from an existing spacetime with asymptotically flat or Anti-de Sitter regions. These wormholes are distinguished by the presence of a local maximum of the scale factor, which allows the analytically continued Lorentzian spacetime to expand after materialization. We present explicit numerical wormhole solutions supported either by an axionic field or a magnetic gauge field, in both cases in conjunction with a self-interacting scalar field. More exotic solutions, with multiple extrema of the scale factor, are also described. As we discovered recently, in the limit of small axionic or magnetic charge, wineglass wormhole solutions split into two separate geometries, one being the background spacetime and the other a disconnected no-boundary instanton. We study the associated topology changing transition in detail and provide an extensive discussion of both the properties and puzzles exhibited by this common family of wineglass/no-boundary instantons. 
\end{abstract}
	
	\maketitle
	
	\tableofcontents

\vspace{1cm}
\section{Introduction}

The history of our universe is known with great confidence back to less than a second after the putative big bang singularity, but if we ask what occurred before the start of the hot, dense big bang phase, then our confidence drops abruptly and the possibilities multiply. A leading model for explaining both the large-scale features of the universe and the  fluctuations on all scales in the cosmic microwave background is inflation \cite{Guth:1981ff,Linde:1981mu,Albrecht:1982wi}. But it remains unknown whether an inflaton field with a suitable potential exists, and even if it does (which we will assume here), it remains an open question of how inflation could have started. The question is pertinent, as the conditions for inflation to start are highly non-generic: one requires a Hubble-sized region of the universe over which the scalar field is roughly homogeneous, sits suitably high up on its potential, and has everywhere a small kinetic energy, see e.g. \cite{Ijjas:2013vea}. Moreover, if one follows inflation back classically, one generically reaches a spacetime singularity, which precludes a satisfactory explanation \cite{Farhi:1986ty,Borde:2001nh}.

There are several ways out of this conundrum. For instance, it could be that the classical solution describing the inflationary phase is non-generic -- inflation could have been preceded by a cosmological bounce from a prior, contracting phase of evolution. In this case, the bounce might have acted like a filter, allowing only a highly tuned solution to pass through and give rise to an inflationary phase \cite{Anabalon:2019equ}. Whether such a scenario is viable or not depends sensitively on the physics of the contracting phase, and thus remains highly speculative.

Perhaps a more promising avenue is to consider a quantum origin for the inflationary phase~\cite{Farhi:1989yr}. Here there are two fundamental possibilities: either our universe could have tunneled out of a pre-existing universe (via Coleman-DeLuccia (CdL) instantons \cite{Coleman:1980aw} or wormholes \cite{Giddings:1987cg,Lavrelashvili:1988un}), or could have arisen out of nothing (via no-boundary instantons \cite{Hawking:1981gb,Hartle:1983ai}). The central result of the present work is that, in some settings, these two quantum processes are intimately related and that in fact wormholes and no-boundary instantons form a single family of tunneling solutions.

When contemplating the possibility of tunneling out of a pre-existing spacetime, it makes a significant difference whether this pre-existing spacetime is already inflating itself, or not. If it is, then transitions to a new spacetime bubble are described via CdL instantons and then, evidently, the process can repeat \cite{Aguirre:2006ak}. This leads to the scenario of eternal inflation, which is plagued by infinities and may consequently be unphysical \cite{Rudelius:2019cfh,Jonas:2021xkx}. 

Here, we will rather be interested in the case in which the pre-existing spacetime is either flat space or Anti-de Sitter (AdS) space. This setting is suggested by string theory, in particular the AdS/CFT correspondence \cite{Maldacena:1997re}, which proposes a non-perturbative definition of quantum gravity in terms of spacetimes that approach AdS asymptotically. In this context it becomes pertinent to ask whether it is possible to up-tunnel from the AdS space to an inflationary phase. This question has generated a significant amount of attention recently \cite{Betzios:2024oli,Lan:2024gnv,Betzios:2024zhf,Betzios:2026rbv,Lavrelashvili:2026zsw}.

\begin{figure}
\includegraphics[width=0.7\textwidth]{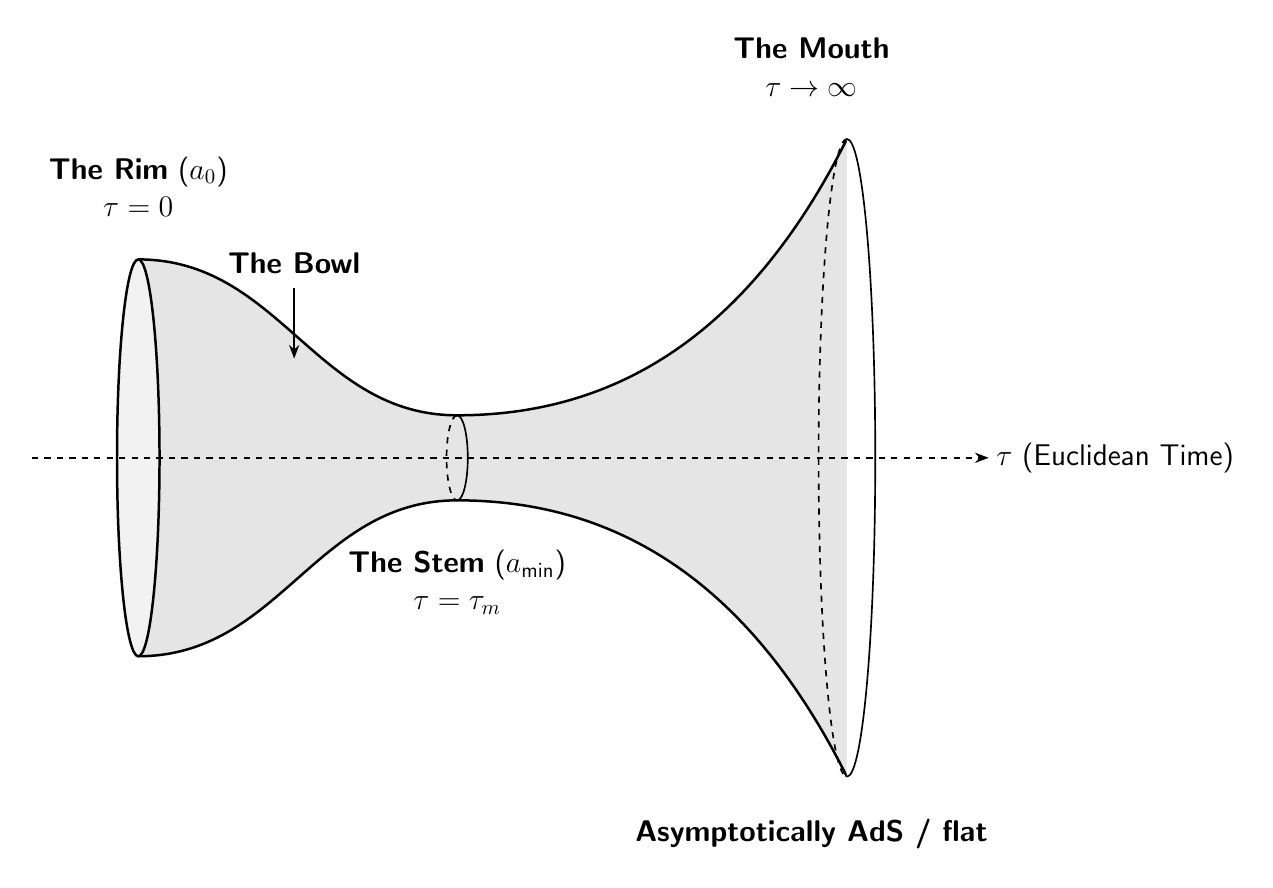}
\caption{Euclidean wineglass wormholes describe the nucleation of an expanding universe (analytically continued on the left) from an asymptotic region that could be flat or AdS space (on the right). These wormholes are distinguished by having both a local minimum of the scale factor (the stem) and a local maximum (the rim), where the latter is necessary to ensure that the resulting baby universe expands rather than crunches.} 
\label{fig:wgw}
\end{figure}

Early Euclidean wormhole solutions, e.g. the axion-supported Giddings-Strominger wormholes \cite{Giddings:1987cg}, described the nucleation of a baby universe out of a pre-existing flat spacetime. However, upon analytic continuation of the wormhole to Lorentzian time, these wormholes lead to contracting universes that crunch in a short timespan and therefore do not constitute realistic tunneling channels for explaining the beginning stage of our universe. In this context, it was shown in \cite{Lavrelashvili:1988un} that the inclusion of an additional self-interacting scalar field significantly alters the geometry of the wormhole throat and enables the nucleation of expanding universes because the wormhole ends at a local maximum, rather than a minimum, of the scale factor (note the simple fact that the function $a_0 - \frac{a_2}{2}\tau^2$ with Euclidean time $\tau$ analytically continues to $a_0 + \frac{a_2}{2} t^2,$ where $t=-i\tau$ denotes the Lorentzian time, and with $a_{0,2}>0$ being constants). Through this mechanism, Euclidean axion ``wineglass'' wormholes were introduced \cite{Lavrelashvili:1988un} and demonstrated to serve as candidates for the beginning of an inflationary epoch -- see Fig.~\ref{fig:wgw} for a sketch of the geometry involved. This framework was further explored in \cite{Rubakov:1988wx}, and asymptotically flat wineglass wormholes have recently been studied in detail using up-to-date computational methods \cite{Jonas:2023ipa}. The addition of a self-interacting scalar to the Giddings-Strominger setup provides a crucial bridge between Euclidean wormhole physics and the standard inflationary paradigm.

An alternative quantum channel for the nucleation of the universe is the no-boundary proposal, in which the universe is created out of nothing. The idea of Hartle and Hawking is that the geometry of the universe, rather than starting from a singularity, should be rounded off, by fiat preventing a singularity and enabling space and time to be entirely self-contained \cite{Hartle:1983ai}, see \cite{Lehners:2023yrj} for a review. No-boundary solutions require a positive vacuum energy (otherwise the geometry cannot be smoothly rounded off), so that they quite naturally provide initial conditions that are suitable for a subsequent inflationary period.

\begin{figure}
\includegraphics[width=0.7\textwidth]{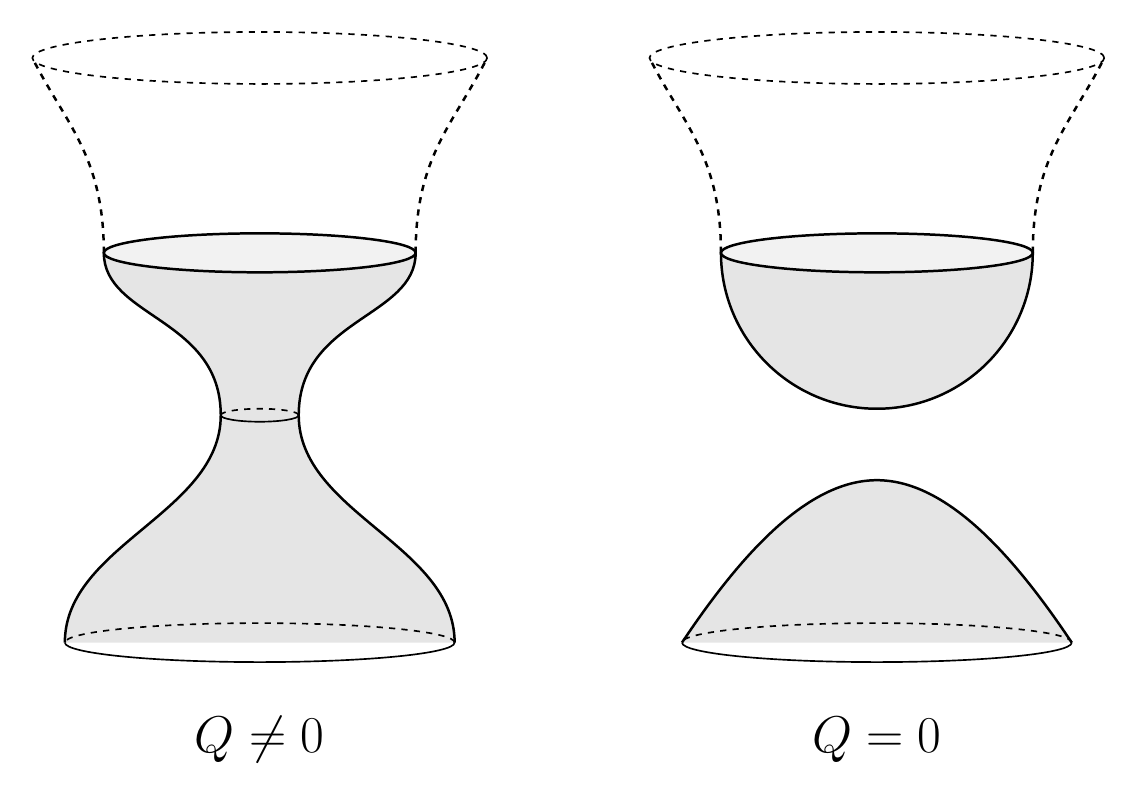}
\caption{As the charge $Q$ of the wineglass wormhole is reduced, the stem becomes thinner and thinner (left panel). In the limit of zero charge, a topological transition takes place: the stem pinches off, leaving behind a no-boundary instanton and a disconnected flat/AdS space (right panel).} 
\label{fig:toptrans}
\end{figure}

A priori, the processes of creation of an inflating universe from nothing or creation via wormholes appear to be entirely distinct. However, as we discovered recently \cite{Lavrelashvili:2026zsw}, they are in fact closely related: wormhole solutions are supported by a charge $Q$, which we will take to be provided either by an axion field or a magnetic gauge field. Moreover, the scalar field interpolates between an asymptotic region and a positive region of the scalar potential, where inflation is meant to take place after nucleation. When the charge $Q$ is reduced, the stem of the wormhole becomes thinner, see Fig.~\ref{fig:toptrans}. Also, the stem effectively separates the part of the solution where the scalar field resides in the inflationary region from the part where the scalar reaches its asymptotic value. In the limit where the charge goes to zero, the stem reaches zero radius, and the bowl of the wineglass geometry separates from the asymptotic region. This is a topological transition which leaves behind a smooth no-boundary geometry, as well as a disconnected asymptotic flat/AdS region.

Thus we find that wineglass wormholes and no-boundary instantons are really part of a common family of tunneling solutions, parameterized by the charge $Q.$ In the present paper, our aim is to argue for the universality of this result. We will consider both axionic and magnetic charges, and both flat (section \ref{sec:flat}) and asymptotically AdS spaces (section \ref{sec:ads}). We will also describe the existence of a number of more exotic solutions, containing multiple stems and/or interpolating multiple times across potential barriers (section \ref{sec:exotic}). Of particular interest is the zero charge limit, which will be described in detail in section \ref{sec:zero}. Section \ref{sec:discussion} provides an extensive discussion not just of the features of these solutions, but mainly of the puzzles that their existence presents. We also include an appendix, reviewing with the example of a $SU(2)$ gauge theory how magnetic solutions can provide the required charges for supporting wormholes. 


\section{Model and equations of motion}


We are searching for Euclidean wormhole solutions that can describe the nucleation of an expanding baby universe, potentially able to generate a new region of spacetime with conditions appropriate for a subsequent inflationary phase. Even though inflation, if it occurred, must have taken place at a rather high energy scale, the absence of primordial B-mode polarization in the cosmic microwave background bounds the Hubble rate to  $5$ orders of magnitude below the Planck scale \cite{Planck:2018jri}. At such an energy/curvature scale, we do not expect higher-order gravitational corrections to be important, and hence we will take the gravitational part of the theory to be given simply by the Einstein-Hilbert action, possibly with suitable boundary terms, which we will discuss. We will be interested in Euclidean Robertson-Walker-type metrics containing a $3$-sphere with line element $\mathrm{d}\Omega_3^2,$
\begin{align}
\mathrm{d}s^2 = \mathrm{d}\tau^2 + a^2(\tau)\mathrm{d}\Omega_3^2\,,
\end{align}
where $\tau$ denotes the Euclidean time and $a(\tau)$ the scale factor. The reason for focusing on such highly symmetric metrics is that one expects these to dominate over less symmetric configurations in the gravitational path integral.

We will consider two types of matter that can support a wormhole solution, either an axion \cite{Giddings:1987cg} or radiation for which the magnetic component is larger than the electric one \cite{Hosoya:1989zn,Gupta:1989bs}. For consistency, the axionic charges $Q_a$ or magnetic charges $Q_m$ must also be distributed in a spherically symmetric manner. In a previous paper  \cite{Jonas:2023ipa} we reviewed in detail how this can be done in the axionic case. In appendix \ref{sec:appendix} we review an example of how this can be done for a $SU(2)$ gauge theory allowing for a magnetic solution. In both cases, we will restrict to configurations with constant charges~$Q_{a,m}$.

As remarked in the introduction, this setup is not quite sufficient yet: in order to obtain wormholes that lead to an expanding, rather than a contracting, universe upon analytic continuation to Lorentzian time, one also requires a scalar field $\phi(\tau)$ with a potential $V(\phi)$ \cite{Lavrelashvili:1988un}. Putting it all together, and specializing to our metric ansatz, the action thus reads \cite{Jonas:2023ipa,Lavrelashvili:2026zsw}
\begin{align}
\frac{S_E}{2\pi^2} = & \int \!\mathrm{d}\tau \left( -3 a\dot{a}^2 + \frac{1}{2}a^3 \dot\phi^2 - 3a + a^3 V(\phi) + \frac{Q_a^2}{a^3} + \frac{Q_m^2}{a}\right) \nonumber \\ & + c_f 3 a^2 \dot{a}\mid_{\tau=\tau_f} \, - \, c_0 3 a^2 \dot{a}\mid_{\tau=0}\, + S_{ct}\,, \label{action}
\end{align}
where $\tau$ derivatives are indicated by dots. So far, we left open the question of whether we should add the Gibbons-Hawking-York boundary term, and thus impose Dirichlet boundary conditions (i.e. fix the boundary value of the scale factor), or not, corresponding to Neumann boundary conditions (i.e. fixing the derivative of the scale factor). We assume that the Euclidean time runs from $\tau=0$ to $\tau=\tau_f.$ The choices of boundary conditions are then encapsulated in the coefficients above, with $c_{0,f}=1$ corresponding to Neumann and $c_{0,f}=0$ corresponding to Dirichlet conditions. The distinction between the two conditions will be important in order to obtain the correct value of the action. We also allow for counter terms $S_{ct},$ which will be required to cancel divergences in the action when the asymptotic geometry of the wormhole solutions is given by Anti-de Sitter (AdS) space.

The equations of motion following from the action are
\begin{align}
\ddot\phi + 3 \frac{\dot{a}}{a}\dot\phi - V_{,\phi} = 0 &\,, \label{scalareom} \\
 3\frac{\ddot{a}}{a} + \dot\phi^2 + V - 2\frac{Q_a^2}{a^6} - \frac{Q_m^2}{a^4} = 0 & \,,\label{acceleration}
\end{align}
while the associated constraint is given by
\begin{align}
3 \frac{\dot{a}^2}{a^2} - \frac{3}{a^2} =  \frac{1}{2} \dot\phi^2  - V(\phi) - \frac{Q_a^2}{a^6} - \frac{Q_m^2}{a^4}\,. \label{constraint}
\end{align}
As usual, the acceleration equation \eqref{acceleration} can be obtained from a time derivative of the constraint, and using the scalar equation of motion \eqref{scalareom}.

\begin{figure}
\includegraphics[width=0.55\textwidth]{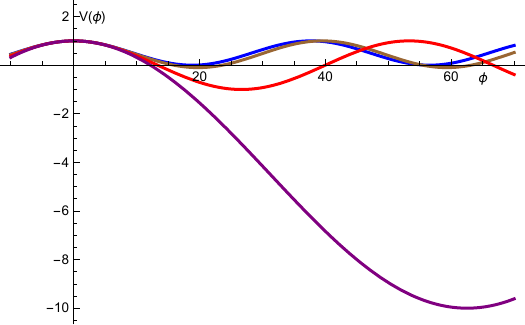}
\caption{We will consider scalar potentials of sinusoidal form, parameterised such that the maxima are at $V=1$ and the minima at $V=V_{min}=(0,-1/10,-1,-10)$ for the blue, brown, red and purple curves respectively. The decay constants are chosen such that the curvature at $\phi=0$ is identical in each case.} \label{fig:pot}
\end{figure}

It remains for us to specify the form of the scalar potential. In the axionic context, it is natural to consider a sinusoidal potential. Moreover, this kind of potential has the characteristics we require, namely a potential maximum (which can be taken to reside at positive values of the potential) around which an inflationary phase could take place, a potential barrier, and a potential minimum which can be freely adjusted, and which can thus allow for a Minkowski or an AdS vacuum. Thus we will consider potentials of the form
\begin{align}
V(\phi) = \frac{1-V_{min}}{2} \cos\left( \frac{\phi}{f}\right) + \frac{1+V_{min}}{2}\,, \label{potential}
\end{align}
where we have chosen units such that the potential maximum is at unity, with the minimum being at $V_{min}.$ We will consider various values of $V_{min}$ below -- see also Fig.~\ref{fig:pot}. The ``decay constant'' $f$ determines the flatness of the potential near the top; in particular, the curvature of the potential at the top (often called the second slow-roll parameter) is given by 
\begin{align}
\frac{V_{,\phi\phi}}{V}(\phi=0) = \frac{V_{min}-1}{2f^2}\,.
\end{align}
The smallness of this parameter provides an estimate of the length of the inflationary phase. We will choose the decay constants such that the curvature at the top remains identical for all potentials -- more explicitly, we will choose 
\begin{align}
f = 6 \sqrt{1-V_{min}}\,,
\end{align}
where the factor of proportionality is chosen so as to be phenomenologically reasonable, with $V_{,\phi\phi}/V(\phi=0)=-1/72$ in each case.


\section{Asymptotically flat case} \label{sec:flat}


We will start by searching for wineglass wormhole solutions that asymptote to flat space. For this purpose, we will use the potential \eqref{potential} with $V_{min}=0$ and $f=6.$ Solutions of this type were first presented in our earlier paper with C. Jonas \cite{Jonas:2023ipa}, using a different scalar potential. Here we will expand on those results and clarify a number of properties of these solutions. 

Our aim is to find half-wormhole solutions interpolating between the origin at $\tau=0$ and a large radius at $\tau=\tau_f,$ where in principle one may take $\tau_f \to \infty$ at the end of the calculation. In our coordinate system flat space corresponds to $a(\tau) =\tau,$ and thus $\dot{a}=1.$ Hence we can impose asymptotically flat conditions by choosing a Neumann boundary condition ($c_f=1$) at the ``outer'' boundary, setting $\dot{a}(\tau_f)=1.$ The action automatically leads to a Dirichlet variational problem for the scalar field, so that we may fix its asymptotic value to reside at a minimum of the potential, say at $\phi_f = -f \pi.$ The origin at $\tau=0$ will correspond to the rim of the wormhole, cf. Fig.~\ref{fig:wgw}. This is meant to correspond to a local maximum of the scale factor, in order for the analytically continued solution to expand, as we explained in the introduction. But we do not know {\it a priori} what the size of the rim will be. Hence we will also impose a Neumann boundary condition at the origin ($c_0=1$) with $\dot{a}(0)=0.$ Likewise, we would like to set the initial derivative of the scalar field to zero, $\dot\phi(\tau=0)=0$, since this is required for the analytically continued solution to be real Lorentzian. This is possible, since variation of the action with respect to the scalar field leads to a surface term of the form $a^3 \dot\phi \delta\phi,$ which can either vanish by fixing the value of $\phi,$ which is what we are doing on the outer boundary, or by setting $\dot\phi=0,$ which is our condition on the inner boundary.

The wormhole solutions must satisfy the classical equations of motion, and thus also the constraint \eqref{constraint}. This links the initial values of the scale factor and scalar field according to
\begin{align}
     3= a_0^2 V(\phi_0) + \frac{Q_a^2}{a_0^4} + \frac{Q_m^2}{a_0^2}\,. \label{constraintat0}
\end{align}
In the axionic case, one obtains a cubic equation for $x\equiv a_0^2,$ while the magnetic case gives rise to a quadratic equation. In both cases one chooses the largest positive root, while the smaller positive root would give rise to a Giddings-Strominger-type wormhole instead (starting at a minimum of the scale factor). In order for two positive real roots to exist, the discriminant must be positive in both cases. This leads to the conditions
\begin{align}
    Q_a V(\phi_0) < 2\,, \qquad Q_m^2 V(\phi_0) < \frac{9}{4}\,,
\end{align}
which are necessary, but not sufficient, for the existence of wineglass wormhole solutions. Thus we see that the charges cannot be arbitrarily large (if we desire a prolonged inflationary phase after nucleation), though they can be arbitrarily small.

\begin{table}[h]
\centering

\begin{minipage}{0.48\textwidth}
\centering

\begin{tabular}{|c|c|c|}
\hline
$Q_a$ & $\phi_0$ & $a_0$ \\ \hline \hline
1.25 & 9.052082628 & 2.356676759 \\ \hline
1 & 8.342666688 & 2.240568681 \\ \hline
0.75 & 7.382925120 & 2.110928711\\ \hline
0.5 & 5.980590014 & 1.966418719\\ \hline
0.25 & 3.707343635 & 1.816389885\\ \hline
0.1 & 1.641184416 & 1.748064726 \\ \hline
0.05 & 0.8347777193 & 1.736170575 \\ \hline
0.02 & 0.3354034745 & 1.732714767 \\ \hline
\end{tabular}

\end{minipage}
\hfill
\begin{minipage}{0.48\textwidth}
\centering
   
\begin{tabular}{|c|c|c|}
\hline
$Q_m$ & $\phi_0$ & $a_0$ \\ \hline \hline
1.25 & 6.17106607732 & 1.82770274334 \\ \hline
1 & 4.52058740582 & 1.75953621483 \\ \hline
0.75 & 2.75893112529 & 1.72167828147 \\ \hline
0.5 & 1.25368205428 & 1.71674859045 \\ \hline
0.25 & 0.310921282217 & 1.72656738886 \\ \hline
0.1 & 0.0494855298594 & 1.73110195320 \\ \hline
0.05 & 0.0123606501661 & 1.73181108037 \\ \hline
0.02 & 0.00197720901393 & 1.73201233892 \\ \hline
\end{tabular}
\end{minipage}

\caption{Boundary values at rim for wineglass wormholes in a sinusoidal scalar potential with $V_{min}=0\,,$ $f=\sqrt{6},$ with spherically symmetric axionic/magnetic matter, for various values of the charges. {\it Left:} Axionic case ($10$ significant digits). {\it Right:} Magnetic case ($12$ significant digits).} \label{table:flat}
\end{table}

In order to find explicit solutions, we employ a shooting technique. We start with a trial value $\phi_0$ for the scalar field at $\tau=0,$ which then corresponds to a value for the size of the rim $a_0,$ as just discussed. The scalar equation of motion \eqref{scalareom} shows that the Euclidean theory effectively evolves in an inverted potential $-V.$ Since the scalar field starts at rest, it will initially roll downwards in the inverted potential, or upwards in the potential. This implies that the starting value of the scalar should reside on the other side of a potential barrier, compared to the vacuum in which the scalar is supposed to end up. In our case, this implies that $0<\phi_0<f\pi.$ If the scalar starts too close to the top of the potential (which is at $\phi=0$), then the field will turn around before reaching the potential minimum at $\phi=-f\pi.$ This is an undershoot solution. On the contrary, if the scalar starts too far from the barrier, it will gain a large kinetic energy and roll over the potential minimum. This is an overshoot solution. We can use this property to fine-tune the initial value of $\phi_0,$ such that asymptotically the scalar remains in the flat space vacuum for as long as we desire, and up to the level of precision that we desire.

\begin{figure}
\includegraphics[width=0.4\textwidth]{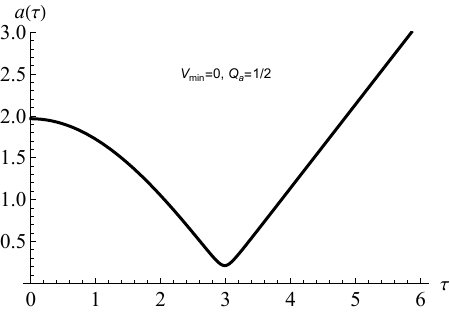}
\hspace{1cm}
\includegraphics[width=0.4\textwidth]{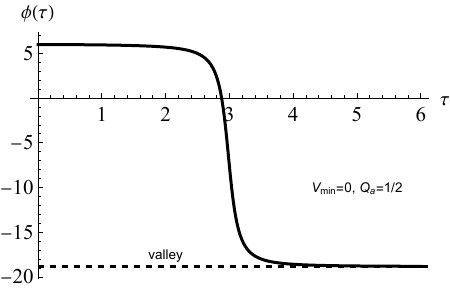}
\caption{An axionic wineglass wormhole with asymptotically flat boundary conditions. {\it Left:} Evolution of the scale factor.  
{\it Right:} Evolution of the scalar field.} \label{fig:V0ax}
\end{figure}

\begin{figure}
\includegraphics[width=0.4\textwidth]{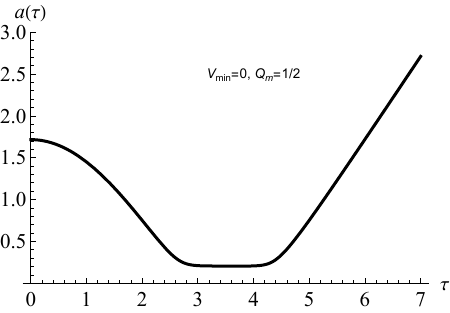}
\hspace{1cm}
\includegraphics[width=0.4\textwidth]{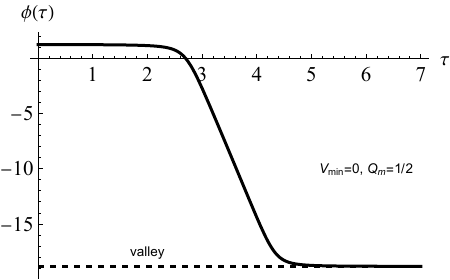}
\caption{A magnetic wineglass wormhole with asymptotically flat boundary conditions. {\it Left:} Evolution of the scale factor.  
{\it Right:} Evolution of the scalar field. } \label{fig:V0mag}
\end{figure}

Table \ref{table:flat} provides the initial conditions for both axionic and magnetic solutions, up to $10$ and $12$ significant digits respectively, for wineglass wormhole solutions with different values of the charges (we only consider solutions with either axionic or magnetic charge, not both simultaneously). Examples of solutions with charges $Q_{a,m}=1/2$ are shown in Figs.~\ref{fig:V0ax} and \ref{fig:V0mag} respectively. As one can see from the figures, the scale factor indeed starts at a local maximum, then contracts and smoothly bounces, before shooting off to infinity in a linear fashion, corresponding to asymptotic flat space. Meanwhile, the scalar field starts near the top of the potential, then crosses the top around the time of the bounce (i.e. in the region of the stem) and approaches the potential minimum at $\phi=-f\pi$ at late ``times''. A distinguishing feature of the magnetic solutions is that the stem region is enlarged compared to the axionic case. In other words, the bowl and the base are further separated, and correspondingly the scalar field takes longer to interpolate between inflationary and asymptotic values.

\begin{figure}
\includegraphics[width=0.45\textwidth]{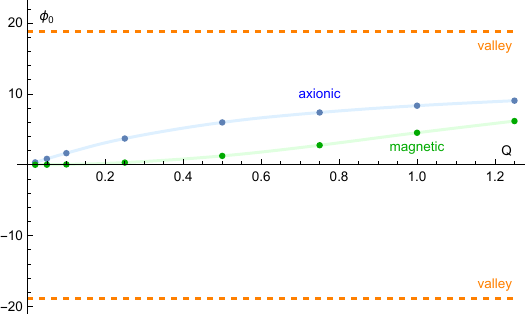} 
\hspace{1cm}
\includegraphics[width=0.45\textwidth]{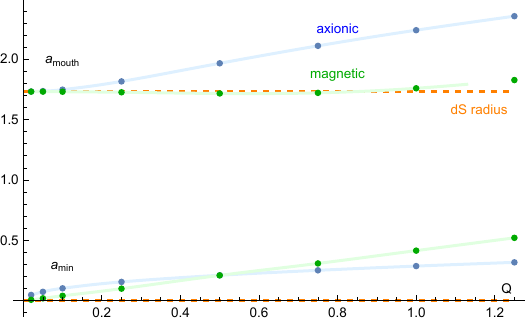}
\caption{{\it Left:} Initial scalar field values for the solutions listed in table \ref{table:flat}. {\it Right:} Sizes of the rim and stem for the same solutions.} \label{fig:V0phi0}
\end{figure}

The initial scalar field values are plotted in the left panel of Fig.~\ref{fig:V0phi0} as a function of the charge. As one can see there, when the charge is smaller, the starting point is closer to the top of the barrier. This implies that solutions with smaller charge lead to analytically continued universes with a longer inflationary phase.

The right panel of Fig.~\ref{fig:V0phi0} shows the sizes of the rim and stem for the same solutions. As one can see, the rim is given to a good approximation (especially when the charges are small) by the de Sitter radius associated with the height of the potential barrier. Thus, when these wormholes nucleate a baby universe, that universe will automatically have a size that is a little larger than the Hubble radius, which fits well with their desired function as providing initial conditions for inflation. The stem is significantly thinner, and tends to zero in the limit of vanishing charge. This central property will be analyzed in detail in section \ref{sec:zero}.

\begin{figure}
\includegraphics[width=0.6\textwidth]{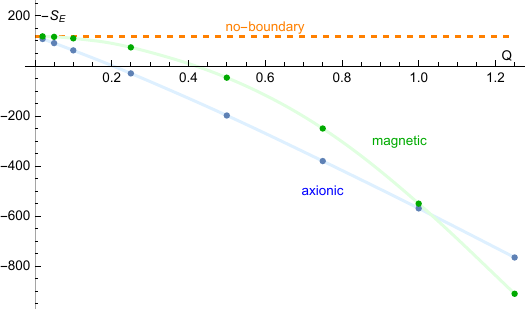}
\caption{Weightings of axionic and magnetic wineglass wormholes with asymptotically flat boundary conditions, for the solutions listed in table \ref{table:flat}.} \label{fig:V0weight}
\end{figure}

It is also of interest to calculate the Euclidean action of these solutions, as the action determines the relative contributions of these various solutions to the gravitational path integral. With the double-Neumann boundary conditions on the scale factor, and making use of the constraint~\eqref{constraint}, the calculation of the on-shell action simplifies and \eqref{action} reduces to
\begin{align}
S^{on-shell}_{NN} = 2\pi^2 \int \mathrm{d}\tau \left(2\frac{Q_a^2}{a^3} + 2\frac{Q_m^2}{a} - a^3 V  \right)\,. \label{actionNN}
\end{align}
Since the asymptotic space will be flat, we do not need to add counter terms to this action. In fact, the contribution of each wormhole solution to the wave function is proportional to minus the Euclidean action,
\begin{align}
    \Psi \sim \sum e^{-S_E}\,, \qquad |\Psi|^2 \sim \sum e^{-2S_E}\,,
\end{align}
up to an overall normalization factor that we will not be concerned with here (though we note that the normalization factor has recently been claimed to lead to highly non-trivial results \cite{Abdalla:2026mxn}, and should hence be studied in future work). Thus we find it more appropriate to plot the weighting, i.e. minus the Euclidean action, and this is shown in Fig.~\ref{fig:V0weight}. This graph exhibits some rather striking features. 

First of all, it shows that the weighting is larger for smaller charges. This means that wormholes with smaller charges come out as preferred. As we saw above, these are also the solutions that lead to a longer inflationary phase. Thus wineglass wormholes preferentially give rise to universes with a lot of inflation (the preference for small charge solutions was previously noticed in \cite{Jonas:2023ipa} and the fact that this leads to longer inflationary phases was the motivating reason behind \cite{Betzios:2024oli}).

Second, even though the weighting of axionic solutions scales approximately linearly with the charge, while that of magnetic solutions scales approximately quadratically, both tend to the same limit in the limit of vanishing charges. Moreover, this limit is far from arbitrary -- here we simply observe that it corresponds to the weighting of a no-boundary instanton sitting at the top of the potential barrier. In the present case this weighting is given by
\begin{align}
    -S_E^{no-bdy}(\phi_0=0) = \frac{12\pi^2}{V_{top}} = 12\pi^2 \approx 118.435\,.
\end{align}
As previously reported in \cite{Lavrelashvili:2026zsw} and as will discuss in detail in section \ref{sec:zero}, this is no coincidence. It is the first hint that wineglass wormholes are closely related to no-boundary instantons.

And third, we notice that the weighting can become positive, corresponding to a negative Euclidean action. This property was already noticed a long time ago in \cite{Shvedov:1996hb}. By inspecting the expression for the on-shell action in \eqref{actionNN}, we see that this can indeed occur quite easily when the charges are small and there is a significant region over which the potential is positive. As we can see from Fig.~\ref{fig:V0weight}, the weighting does however not become arbitrarily positive, as it is bounded by the no-boundary value at the top of the potential. We will return to this observation in section~\ref{sec:zero}, but first turn our attention to wineglass wormholes with AdS asymptotics.


\section{Asymptotically AdS case} \label{sec:ads}


String theory suggests that a possible framework for quantum gravity consists of having a spacetime that is fixed asymptotically. This construction requires the presence of negative vacuum energy, such that the asymptotic spacetime is (locally) AdS space. A pertinent question is then whether a universe like ours, with positive rather than negative vacuum energy and with an expanding, rather than asymptotically fixed geometry, can arise from this setting. An obvious possibility might be that a universe such as ours could form in the interior (bulk), perhaps via a quantum tunneling event. This is precisely the scenario envisaged in \cite{Betzios:2024oli}, making use of wineglass wormholes interpolating between AdS asymptotics and an expanding baby universe. In a previous paper \cite{Lavrelashvili:2026zsw}, we presented the first explicit (numerical) solutions of this type (the solutions were not explicitly constructed in \cite{Betzios:2024oli}). Here we will both simplify and expand our previous results by considering several potentials, all of sinusoidal form.

The potentials we will consider are all of the functional form \eqref{potential}, with $V_{min}=-1/10, -1, -10$ and corresponding values $f=6\sqrt{11/10}, 6\sqrt{2}, 6\sqrt{11}$ respectively. This ensures that near the origin at $\phi=0,$ all potentials have the same height and curvature. In principle, one can easily scale the overall magnitude of the potentials. The reason why we are taking the potentials to have the same inflationary region is that we are considering the following (``top down'' \cite{Hawking:2002af}) question: if we assume that the large-scale properties of the universe, as well as the fluctuations in the cosmic microwave background, all result from an inflationary phase, then what are the probabilities that this could have arisen from various possible previous phases, in particular from various possible AdS vacua?

The strategy for finding solutions will be very similar to that in the flat case, with one notable exception: in a holographic setting, it is natural to fix the metric, rather than its derivative, on the outer boundary.  Thus we will impose a Dirichlet condition, setting $c_f=0$ at $\tau_f$ in \eqref{action}, with the scale factor taking a large, arbitrary value there (and the scalar field approaching the AdS minimum at $\phi_f=-f\pi$). The idea is once again that the boundary will be taken to infinity at the end of the calculation, which poses no problem in practice, as the wormhole solutions we will consider rapidly approach the Euclidean AdS (EAdS) geometry at large $\tau.$ Except for this proviso, all other conditions remain unchanged and the shooting technique used in optimizing initial conditions also remains unchanged. Table \ref{table:ads} shows the optimized initial values for the scale factor and scalar field in all three potentials, for different values of the axionic charge (in the AdS case we only consider axionic solutions for simplicity, though in principle one can straightforwardly extend our techniques to the magnetic case).

\begin{table}[h]
\centering
\begin{tabular}{|c||c|c|c|c|c|c|}
\hline
& \multicolumn{2}{c|}{$V_{min} = -0.1\,, f = 6\sqrt{1.1}$} & \multicolumn{2}{c|}{$V_{min}=-1\,, f=6\sqrt{2}$} & \multicolumn{2}{c|}{$V_{min}=-10\,, f=6\sqrt{11}$} \\
\hline\hline
$Q_a$ & $\phi_0$ & $a_0$ & $\phi_0$& $a_0$ & $\phi_0$ & $a_0$ \\ \hline \hline
1 & 8.1813654 &  2.2262896 & 7.6305928 & 2.1795752 & 7.2442863 & 2.1488023 \\ \hline
0.75 & 7.2672855 & 2.1023588 & 6.8527397 & 2.0726746 & 6.5478672 & 2.0519981 \\ \hline
0.5 & 5.91583169 & 1.9627763 & 5.6688637 & 1.9492412 & 5.4748308 & 1.9390854 \\ \hline
0.25 & 3.6921176 & 1.8159192 & 3.6294470 & 1.8140269 & 3.5758147 & 1.8124918 \\ \hline
0.1 & 1.6416899 & 1.7480842 & 1.6443367 & 1.7481802 & 1.6478378 & 1.7482942 \\ \hline
0.05 & 0.83587900 & 1.7361823 & 0.84098238 & 1.7362368 & 0.84640473 & 1.7362951 \\ \hline
0.02 & 0.33597060 & 1.7327170 & 0.33857457 & 1.7327277 & 0.34127792 & 1.7327388 \\ \hline
\end{tabular}
\caption{Initial values at the rim, for wineglass wormholes in a sinusoidal scalar potential with $V_{top}=1$ and various values for $V_{min},$ as indicated. Here we assume spherically symmetric axionic matter with the charges listed in the table. We provide $8$ significant digits for the rim values of the scalar field and scale factor.} \label{table:ads}
\end{table}

\begin{figure}
\includegraphics[width=0.4\textwidth]{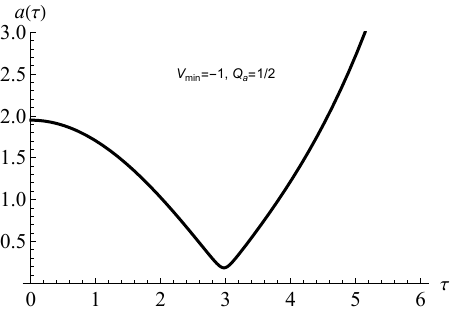}
\hspace{1cm}
\includegraphics[width=0.4\textwidth]{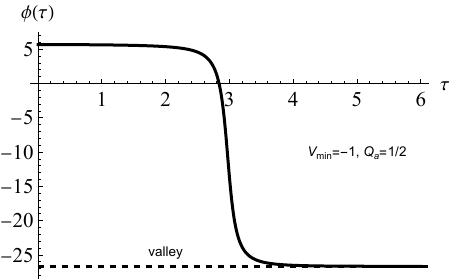}
\caption{An axionic wineglass wormhole with asymptotically AdS boundary conditions. {\it Left:} Evolution of the scale factor.  
{\it Right:} Evolution of the scalar field.} \label{fig:Vminus1sol}
\end{figure}

An example of a wineglass wormhole solution with AdS asymptotics is shown in Fig.~\ref{fig:Vminus1sol}. In this example $V_{min}=-1$ and $Q_a=1/2.$ In the bowl region, the solution is very similar to the corresponding solution with flat asymptotics depicted in Fig.~\ref{fig:V0ax}. The base (wormhole mouth) is of course different, as in the present case the scalar field reaches the AdS minimum at $\phi_f = -6\sqrt{2}\pi$ while the scale factor expands exponentially, with $a(\tau) \propto e^{\sqrt{-V_{min}/3}\, \tau}$ to leading order.

As one can see from table \ref{table:ads}, the scalar field values at the rim mainly vary with the charge, and not so much with the depth of the potential, especially as the charges get small. Here it is again the case that small-charge solutions lead to long inflationary phases upon analytic continuation at the rim, since the scalar starts its Lorentzian evolution at rest and high up on the potential. 

It is of interest once again to also look at the action of these solutions. The expression for the on-shell action differs from the asymptotically flat case in two important aspects. The first was already discussed above, namely that we impose a Dirichlet condition on the scale factor at the outer boundary. With this choice and upon using the constraint \eqref{constraint}, the action \eqref{action} reduces to
\begin{align}
S^{on-shell}_{ND} = 4\pi^2 \int \mathrm{d}\tau \left(\frac{Q_a^2}{a^3} + \frac{Q_m^2}{a} + a^3 V - 3a \right)\, + S_{ct}.
\label{actionND}
\end{align}
The second difference is that, with asymptotically EAdS boundary conditions, the on-shell action diverges at large $\tau$ because the scale factor keeps contributing (exponentially). This divergence must be canceled by the addition of counter terms, which renormalize the action \cite{Balasubramanian:1998sn}. When the AdS vacuum sits at a local maximum of the potential, this is a rather involved procedure, as the scalar field also contributes diverging terms at large volume, and these divergences must be renormalized in order to obtain physically sensible results, see e.g. \cite{deBoer:1999tgo,Elvang:2016tzz}. We analyzed a case of this sort in our previous paper, and refer to this discussion for details \cite{Lavrelashvili:2026zsw}. Here, however, we are in a much simpler situation, as the AdS vacua we are interested in are local minima of the potential. This implies that the scalar field converges rapidly to its asymptotic value, and does not cause divergences in the action (see equation (S7) in \cite{Lavrelashvili:2026zsw} and the associated discussion). The counter terms are thus purely gravitational, and are given by the standard expression \cite{Balasubramanian:1998sn}
\begin{align}
    S_{ct} & = \frac{1}{2} \int_{outer} \mathrm{d}^3x \sqrt{h} \left( \frac{4}{L} + L R^{(3)}\right)\\ &= 2\pi^2 \left( \frac{2}{L} a^3 + 3La\right)\mid_{\tau=\tau_f}
\end{align}
where $L=\sqrt{-3/V_{min}}$ is the AdS radius of curvature.

\begin{figure}
\includegraphics[width=0.6\textwidth]{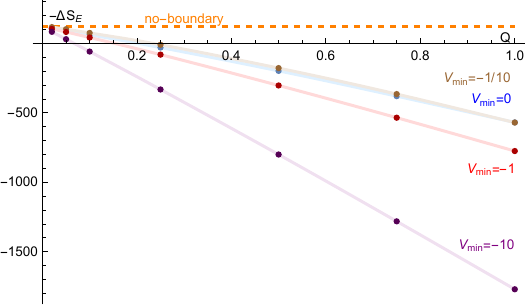}
\caption{``All roads lead to Rome.'' Renormalized, background subtracted weightings of axionic wineglass wormholes with various asymptotically AdS boundary conditions. At small values of the charge, all weightings tend to a common value.} \label{fig:adsweight}
\end{figure}

Since we are interested in relative probabilities, we again plot the weighting (i.e. the negative of the Euclidean action) in Fig.~\ref{fig:adsweight}. In fact, since we start from different AdS backgrounds with non-trivial actions themselves, the physically sensible quantity to consider is rather the background-subtracted weighting,
\begin{align}
    -\Delta S_E = - \left(S^{on-shell}_{ND} - S_{EAdS}\right) = - \left(S^{on-shell}_{ND} + \frac{12\pi^2}{V_{min}}\right)\,,
\end{align}
and this is the quantity shown in Fig.~\ref{fig:adsweight}. The first thing to note is that, just as in the flat case, the weightings in a given potential scale approximately linearly with the axionic charge $Q_a.$ Also, we can see that small charge solutions are once again preferred over large charge wormholes. This confirms that, also with AdS asymptotics, wineglass wormholes preferentially nucleate universes with longer inflationary phases. 

If we now compare the different potentials to each other, we can see that deeper potential minima are disfavored and shallower ones preferred. Nucleating a baby universe from a GUT or Planck-scale minimum is thus much less likely than originating from a much smaller energy-scale AdS vacuum, although we can also see that the weightings start clustering together at small values of the AdS depth -- note that the weightings for $V_{min}=-1/10$ lie almost exactly on top of those for asymptotically flat solutions with $V_{min}=0$ (which are included in the plot for comparison).

However, and this is surely the main point, in the parameter region where the most likely solutions lie, namely at small charges, the differences between potentials shrink away, and in fact all weightings tend to the same value in the limit of vanishing charge. This limit can once again be identified with the weighting of the no-boundary instanton sitting at the top of the potential barrier. Thus, overall, the wormholes that contribute most to the gravitational path integral are those with the smallest possible charges, irrespective of which potential depth they interpolate from. And these are also the wormholes giving rise to the longest inflationary phases. 

But before discussing the vanishing-charge limit further, we wish to show that in addition to the ``standard'' wineglass wormholes exhibited so far, the theories under consideration here also admit more exotic solutions.

\section{Exotic solutions} \label{sec:exotic}

Even though our Robertson-Walker-type ansatz for the metric and matter fields is quite restrictive, since it imposes such a high degree of symmetry, the theories we consider nevertheless admit a broad range of solutions. We will illustrate this with several solutions that may also be termed wormholes, but that admit a more involved profile for both the scale factor and the scalar field. All solutions considered here are characterized by having an asymptotic region where EAdS space is reached, and on the other end they reach a rim that is a local maximum of the scale factor. Thus, they satisfy the same boundary conditions as the solutions considered above. However, they differ by what happens in between the two boundaries. (We exhibited a number of exotic, oscillating solutions with flat asymptotics in \cite{Jonas:2023ipa}.)

\begin{figure}
\includegraphics[width=0.4\textwidth]{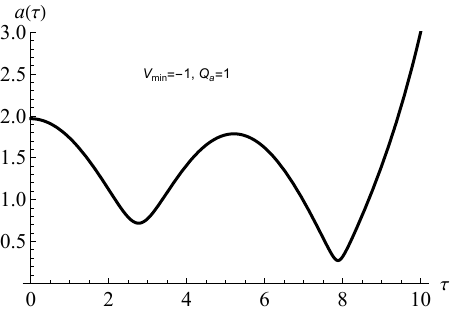}
\hspace{1cm}
\includegraphics[width=0.4\textwidth]{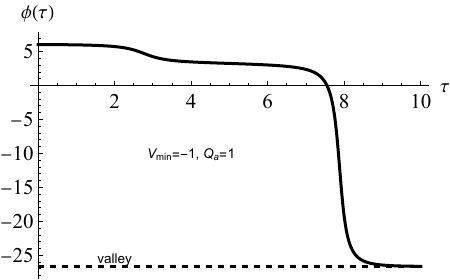}
\caption{A wineglass wormhole with $2$ stems, with $V_{min}=-1,\, Q_a=1,\,$ $\phi_0 = 6.02621345,\,$ $a_0=1.96683280.$
{\it Left:} Evolution of the scale factor.  
{\it Right:} Evolution of the scalar field.} \label{fig:2stems1}
\end{figure}

A first example is shown in Fig.~\ref{fig:2stems1}. This solution contains two stems, i.e. two minima of the scale factor. During the first bounce of the scale factor, the scalar field only moves somewhat further up the potential, and then crosses the barrier to reach the desired AdS vacuum during the second bounce of the scale factor. Perhaps surprisingly, compared to the single-stem solution with the same charge, the two-stem solution starts higher up on the potential. It is also less suppressed, with a renormalized, background-subtracted weighting of $-\Delta S_E \approx -567$ (compared to $-\Delta S_E \approx -775$ for the single stem solution).

\begin{figure}
\includegraphics[width=0.4\textwidth]{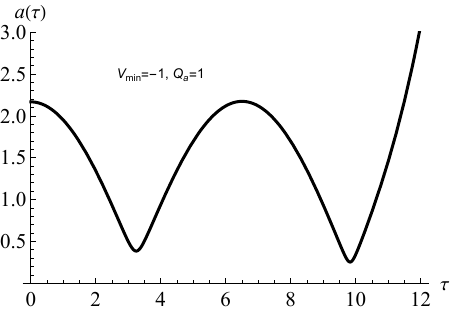}
\hspace{1cm}
\includegraphics[width=0.4\textwidth]{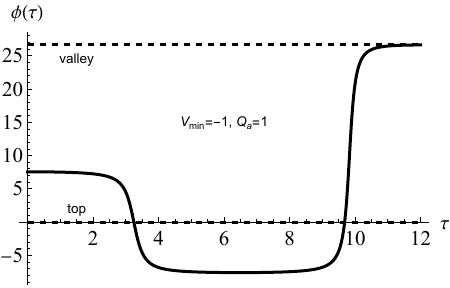}
\caption{A wineglass wormhole with $2$ stems, this time with the scalar interpolating across the potential barrier twice. Here $V_{min}=-1,\, Q_a=1,\, \phi_0 = 7.560478268,\, a_0 = 2.168019383.$
{\it Left:} Evolution of the scale factor.  
{\it Right:} Evolution of the scalar field.} \label{fig:2stems2}
\end{figure}

Before discussing this property further, we would like to point out that a second, inequivalent two-stem solution exists, in which the scalar field crosses the top of the barrier twice, once during each bounce of the scale factor -- see Fig.~\ref{fig:2stems2}. This solution thus ends up in the potential minimum residing at the same side of the barrier as the starting value. For this solution, the scalar field value at the rim is also higher up the potential than for the single-stem solution, but not by much this time. And the background-subtracted weighting is now given by $-\Delta S_E \approx -840,$ so this double-stem wormhole is in fact more suppressed.

\begin{figure}
\includegraphics[width=0.4\textwidth]{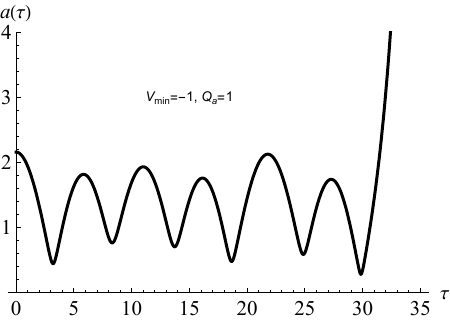}
\hspace{1cm}
\includegraphics[width=0.4\textwidth]{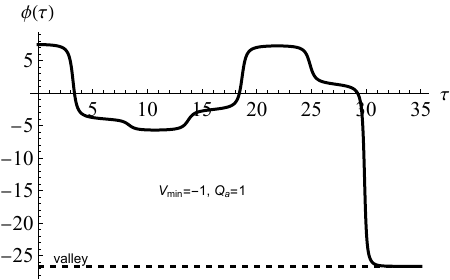}
\caption{A wineglass wormhole with $6$ stems, with $V_{min}=-1, Q_a=1, $ $\phi_0 = 7.4933858,\, $ $a_0 = 2.1571918.$
{\it Left:} Evolution of the scale factor.  
{\it Right:} Evolution of the scalar field.} \label{fig:6stems}
\end{figure}

\begin{figure}
\includegraphics[width=0.4\textwidth]{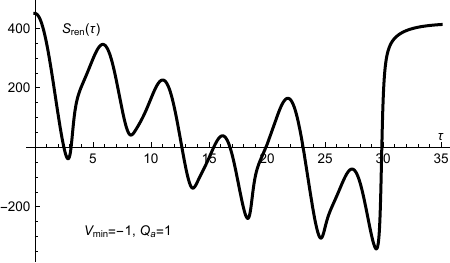}
\caption{The renormalized (but not background-subtracted) Euclidean action integrated from the origin at the rim up to coordinate value $\tau,$ for the solution shown in Fig.~\ref{fig:6stems}. The action gets reduced by each bounce, but the end result is still much more suppressed than the no-boundary limit.} \label{fig:6stemaction}
\end{figure}

But let us get back to the observation that the first double-stem solution had a higher weighting than the single-stem wormhole. This makes one wonder whether solutions with even more bounces of the scale factor might exist, and whether they might in fact be even more preferred. It turns out that such solutions do indeed exist -- see Fig.~\ref{fig:6stems} for an example with $6$ stems. Its weighting is given by $-\Delta S_E \approx -296.$ Hence it is indeed less suppressed. As can be seen from Fig.~\ref{fig:6stemaction}, each bounce with its surrounding quasi-de Sitter geometry reduces the action, but the last bounce mostly compensates for these reductions in the Euclidean action. In fact, even the $6-$stem solution is still highly suppressed compared to the no-boundary limit of small-charge solutions. We were not able to find solutions with more than $6$ bounces, though they might of course exist. Since these multi-stem solutions start higher up on the potential, there is however limited ``room'' for their starting values. What tends to happen is that below a certain threshold value of $\phi_0,$ the scalar field just oscillates around the top of the potential barrier (with the scale factor oscillating many times), without ever reaching an AdS vacuum. Then the desired boundary conditions cannot be satisfied and no further wormhole solution exists.

\begin{figure}
\includegraphics[width=0.4\textwidth]{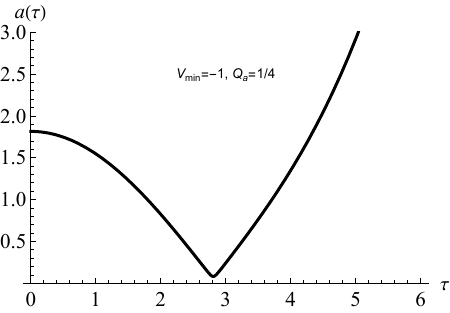}
\hspace{1cm}
\includegraphics[width=0.4\textwidth]{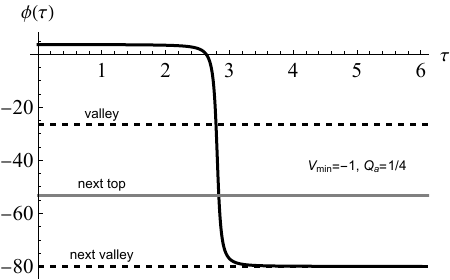}
\caption{A wineglass wormhole interpolating over two adjacent potential barriers. Here $V_{min}=-1,\, Q_a=1/4,\, \phi_0=3.64841323,\, a_0=1.81495778.$ {\it Left:} Evolution of the scale factor.  
{\it Right:} Evolution of the scalar field.} \label{fig:2hills}
\end{figure}

Finally, due to the fact that the potentials we have chosen are periodic, there is the possibility that the scalar field may cross more than one potential barrier. We show an example of such a solution in Fig.~\ref{fig:2hills}. This solution looks very similar to the standard single-stem solutions described above, the only difference being that the scalar field picks up a slightly higher kinetic energy when crossing the potential barrier, and this is sufficient for it to fly over a second barrier and land in the following potential well. This solution has a background-subtracted weighting of $-\Delta S_E \approx -453$ and is significantly more suppressed than the corresponding single-barrier solution with $Q_a=1/4$ and $-\Delta S_E \approx -81.$


\section{No-boundary limit} \label{sec:zero}

Of special interest is the zero-charge limit of the wineglass wormhole solutions we have discussed so far. In our numerical solutions we noticed two effects in this limit: the size of the stem shrinks to zero, cf. Fig.~\ref{fig:V0phi0}, and the weighting tends to that of a no-boundary instanton sitting at the top of the potential barrier, cf. Figs.~\ref{fig:V0weight} and \ref{fig:adsweight}. As first argued in our earlier paper \cite{Lavrelashvili:2026zsw}, this indicates that a topological transition takes place, in which the wormhole solution splits into two pieces, the first being a no-boundary instanton and the second being the background flat or EAdS space.

\begin{figure}
\includegraphics[width=0.4\textwidth]{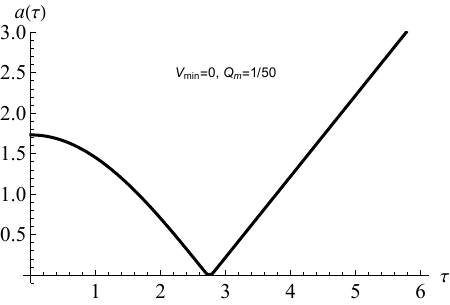}
\hspace{1cm}
\includegraphics[width=0.4\textwidth]{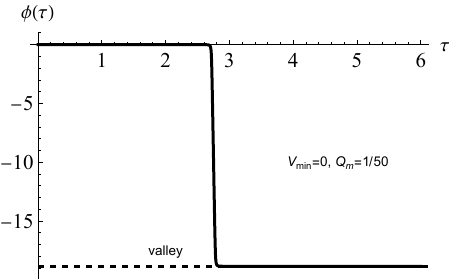}
\caption{Small-charge limit: A magnetic wineglass wormhole with asymptotically flat boundary conditions with $Q_m=1/50$. 
{\it Left:} Evolution of the scale factor.  
{\it Right:} Evolution of the scalar field. 
The evolution consists of a near-dS phase with approximate metric $a(\tau)=\sqrt{3}\sin(\tau/\sqrt{3}+\pi/2)$ up to $\tau_{t}=\sqrt{3}\pi/2,$ followed by the approximately linear flat space solution. The deviation from the approximate solutions is smaller than the width of the curve used to draw this plot.} \label{fig:V0thinmag}
\end{figure}

We can look at this effect in more detail. Fig.~\ref{fig:V0thinmag} shows an example of a magnetically charged wineglass wormhole solution with very small charge, in this case $Q_m=1/50.$ What we can see is that the stem has become very small (cf. the prolonged bounce seen for $Q_m=1/2$ in Fig.~\ref{fig:V0mag}). This is accompanied by a very sharp transition of the scalar field. In fact, the scalar field spends almost all of its time near the extrema of the potential, interpolating abruptly from very close to the top to the AdS minimum, just as the scale factor bounces. In the bowl region, the metric becomes virtually indistinguishable from that of Euclidean de Sitter (EdS) space, i.e. it becomes a hemisphere. One might say that the small-charge limit is also the thin-wall limit of the wormhole solutions.

We can gain a better understanding of this thin-wall limit by analyzing the stem region in somewhat more detail. We can obtain an analytic approximation to the behavior of the scale factor close to the minimum $a_{min}$ by expanding the equations of motion near $a_{min}.$ In the axionic case, this leads to the approximation
\begin{align}
a_{ax}(\tau) = \sqrt{a_{min}^2 + K_2^{ax} (\tau - \tau_{min})^2}\,, \label{stemax}
\end{align}
where the subscript $min$ refers to quantities evaluated at minimum scale factor, and where
\begin{align}
K_2^{ax} = 2 - a_{min}^2 V_{min}\,.
\end{align}
The minimum size of the scale factor can also be obtained as a function of the charge, by solving the constraint \eqref{constraint}, 
\begin{align}
    a_{min} = \frac{Q_a^{1/2}}{3^{1/4}}+\frac{Q_a^{3/2} (V_{min}-\frac{1}{2}\dot\phi^2_{min})}{12\ 3^{3/4}}+O\left(Q_a^{5/2}\right)\,, \qquad \textrm{(axionic case)} \,. \label{stemchargeax}
\end{align}
This confirms that in the zero-charge limit, the stem disappears.

\begin{figure}
\includegraphics[width=0.4\textwidth]{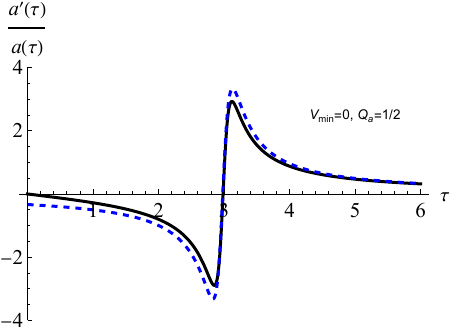}
\hspace{1cm}
\includegraphics[width=0.4\textwidth]{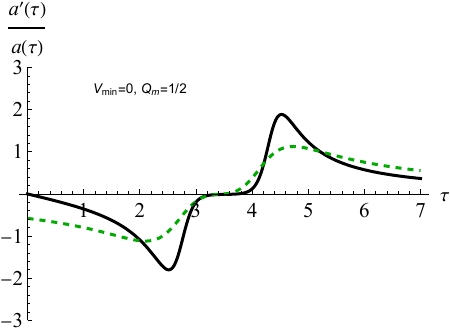}
\caption{Friction term $\dot{a}/a,$ with the solid line corresponding to the numerical solution and the dashed line to the approximation in \eqref{stemax} respectively \eqref{stemmag}, for the axionic wormhole solution shown in Fig.~\ref{fig:V0ax} (left panel) and for the magnetic solution shown in Fig.~\ref{fig:V0mag} (right panel).} \label{fig:fric}
\end{figure}

It is of interest to look at the friction term, $\dot{a}/a,$ that appears in the equation of motion for the scalar, Eq. \eqref{scalareom}. It is shown in the left panel of Fig.~\ref{fig:fric}, for a solution with charge $Q_a=1/2$ (we chose a slightly larger charge for better visibility). The figure shows both the numerical solution and the friction term derived from the approximation \eqref{stemax} above. As one can see, the agreement is very good. The interpretation is as follows: at small $\tau,$ we have anti-friction, which eventually drives the scalar away from its location near the top of the potential barrier. The scalar then moves quickly as the scale factor bounces. Meanwhile, the friction term turns positive and becomes large, thereby slowing the scalar field down and eventually causing it to stop. It then comes to rest at the potential minimum. The approximate expression \eqref{stemax} implies that the peaks in the friction term occur at times $\Delta \tau \sim a_{min}$ away from the bounce of the scale factor. Combining this relation with \eqref{stemchargeax}, we see that the wall indeed gets thinner in proportion to $Q_a^{1/2}.$ This explains why the scalar interpolates so abruptly between the extrema of the potential when the charge is small.

In the magnetic case, because of the comparatively long bounce, it turns out that one needs to expand to higher order to obtain a useful approximate expression. Expanding to fourth order and resumming, we get
\begin{align}
a_{mag}(\tau) = \sqrt{a_{min}^2 + K_2^{mag} (\tau - \tau_{min})^2 + K_3^{mag} (\tau - \tau_{min})^3 + K_4^{mag} (\tau - \tau_{min})^4} ~, \label{stemmag}
\end{align}
where
\begin{align}
K_2^{mag} &= 1 - \frac{1}{6} a_{min}^2 \dot{\phi}_{min}^2 - \frac{2}{3} a_{min}^2 V_{min}\,,  \\
K_3^{mag} &= -\frac{1}{3} a_{min}^2 \dot{\phi}_{min} V'_{min}\,,\\
K_4^{mag} &= \frac{(K^{mag}_2)^2}{4a_{min}^2} + \frac{K^{mag}_2}{12} \left( \frac{7}{6}\dot{\phi}_{min}^2 + \frac{2}{3}V_{min} - \frac{3}{a_{min}^2} \right) - \frac{a_{min}^2}{12} \left( (V'_{min})^2 + V''_{min} \dot{\phi}_{min}^2 \right)\,. 
\end{align}
In this case, one can also find an expression for the size of the stem in terms of the charge, again by solving the constraint \eqref{constraint}. This time we obtain
\begin{align}
    a_{min} = \frac{Q_m}{\sqrt{3}}+\frac{Q_m^3 (V_{min}-\frac{1}{2}\dot\phi_{min}^2)}{18 \sqrt{3}}+O\left(Q_m^5\right)\,, \qquad \textrm{(magnetic case)} \,.\label{stemchargemag}
\end{align}

The friction term for a magnetic solution with $Q_m=1/2$ is shown in the right panel of Fig.~\ref{fig:fric}, again with the solid line being the numerical solution and the dotted line derived from the approximation \eqref{stemmag}. Clearly, the approximation is less good than in the axionic case, even though we expanded to fourth order. This has to do with the fact that the bounce is more prolonged for magnetic solutions, which is a consequence of the fact that the charge term in the equations of motion varies less rapidly compared to the axionic case. Nevertheless, the broad features of the axionic case are reproduced: the friction term first acts as anti-friction and then, after the bounce, becomes a friction term that slows the scalar down again. Also in this case, the width of the wall is roughly given by the stem size, $\Delta\tau \sim a_{min},$ which in turn becomes small as the charge becomes small. Thus, we obtain a very similar thin-wall limit as the charge is reduced.

We can use the expressions above to understand why the wall does not contribute significantly to the action in the thin-wall limit. The argument is very simple: the integrals for the on-shell actions with flat or AdS asymptotics, Eqs. \eqref{actionNN} and \eqref{actionND}, contain only the terms 
\begin{align}
    \frac{Q_a^2}{a^3} &\sim Q_a^{1/2}\,, \quad a^3 V \sim Q_a^{3/2}V_{min}\,, \quad a_{min} \sim Q_a^{1/2}\,, \qquad \textrm{(axionic case)}\,, \\
    \frac{Q_m^2}{a} &\sim Q_m\,, \quad a^3 V \sim Q_m^{3}V_{min}\,, \quad a_{min} \sim Q_m\,, \qquad \textrm{(magnetic case)}\,,
\end{align}
where we have approximated these terms near $a_{min}.$ Moreover, as discussed above, the width of the wall is of order $a_{min},$ i.e. of order $Q_a^{1/2}$ or $Q_m,$ and hence it is clear that the integral over the wall region disappears in the zero-charge limit. This implies that in the zero-charge limit, the action neatly splits into two pieces, one being the EdS bowl (which is a no-boundary instanton) and the background space. The weighting of the no-boundary part thus becomes \cite{Lehners:2023yrj}
\begin{align}
    -S_E^{no-bdy} = \frac{12\pi^2}{V_{top}}\,.
\end{align} 
In the zero-charge limit, the background subtraction scheme still functions, since we may write
\begin{align}
    -\Delta S_E(Q=0) &= -[S_E^{ren}(Q=0)-S_{EAdS}] \nonumber \\ &= -[S_{EAdS}+S_E^{no-bdy}-S_{EAdS}] \nonumber \\ & = -S_E^{no-bdy}\,.
\end{align}
Thus, from the point of view of the action, the topological transition is a smooth process. Of course, for the scalar field it cannot be smooth, as it needs to interpolate between the two vacua, and this interpolation becomes sharper and sharper, effectively turning into a step function when $Q_{a,m}=0$. Correspondingly, it is only when the charges are exactly zero that the geometry can smoothly cap off, as is evident from the constraint Eq. \eqref{constraint} by inspection.

\begin{figure}[htbp]
\centering
\includegraphics[width=0.7\textwidth]{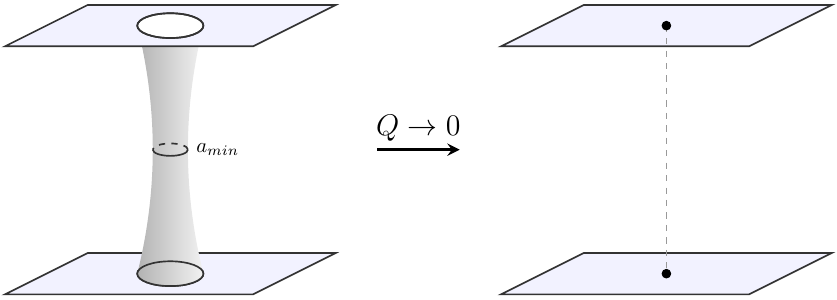}

\vspace{1cm}

\includegraphics[width=0.7\textwidth]{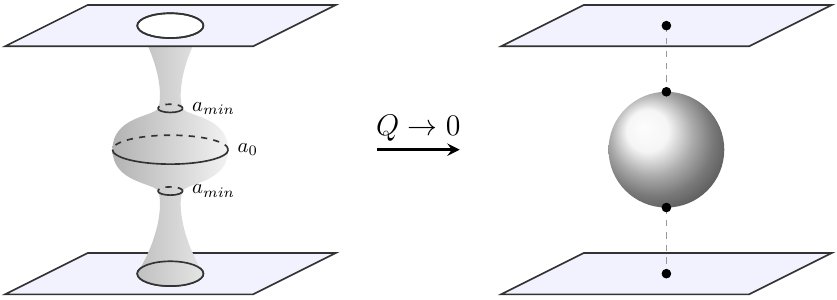}
\caption{The zero-charge limit of full wormholes. {\it Upper panel:} Ordinary wormholes see their throat shrink away in the zero-charge limit, leaving behind two disconnected asymptotic regions. {\it Lower panel:} Full wineglass wormholes by contrast turn not only into two asymptotic regions, but in addition leave behind a Euclidean de Sitter sphere.} \label{fig:zerocharge}
\end{figure}

We can also think about what the zero-charge limit means for Giddings-Strominger-type wormholes. These (half-)wormholes end at a minimum of the scale factor, and as the charge gets reduced, the minimum size shrinks to zero. In the limit of zero charge, the wormhole thus simply disappears, and it leaves behind only the background space. If we think about full wormholes, linking two asymptotic regions of spacetime, then the zero-charge limit would correspond to the disappearance of the wormhole, and would leave behind two disconnected spacetimes, for a sketch see Fig.~\ref{fig:zerocharge}.

By contrast, when considering full wineglass wormholes, also interpolating between two asymptotic regions of spacetime, then the zero-charge limit would leave behind not just the two asymptotic regions, but also a disconnected sphere, corresponding to a full Euclidean de Sitter solution. It would be interesting to see if this effect could have implications for the factorization puzzle in AdS/CFT, see e.g. \cite{Marolf:2021kjc}.


\section{Discussion} \label{sec:discussion}

We have discussed explicit Euclidean wineglass wormhole solutions that can mediate the nucleation of expanding baby universes. These solutions are supported either by an axionic or a magnetic charge, and in addition require the presence of a scalar field with a potential that contains a barrier. As we found, when the charges are small, the scalar field starts its evolution high up on the potential, very close to the top of the barrier, in the analytically continued baby universe. This can provide just the right initial conditions for a long inflationary phase. Moreover, the weighting of these solutions is larger for smaller charges, so that baby universes with longer inflationary phases come out as preferred. 

But the small-charge limit holds a further surprising aspect: as the charge is reduced, the stem of the wormhole solutions shrinks, and in the limit of zero charge pinches off, leading to a topological transition in which the background spacetime separates from the bowl of the wormhole, and the bowl turns into a no-boundary instanton. This instanton has an even higher weighting than the wormhole solutions, and mediates the nucleation of a universe in which the scalar field sits right on top of the barrier in the potential.

Wineglass wormholes and real Euclidean no-boundary instantons may thus be seen as being part of a common family of solutions, parameterized by the axionic or magnetic charge. It is interesting, and characteristic of quantum gravity, that within this family the spacetime topology may change, and this supports the view that different spacetime topologies may contribute to the wave function of the universe. 

For these reasons, solutions belonging to this family may play a distinguished role in explaining the initial conditions for our universe, but their existence also raises a number of conceptual puzzles. In concluding, we will discuss a few such puzzles, which will hopefully be resolved in future work.

One question one may wonder about is whether the charge is part of the theory or part of the solution, i.e. whether it is fixed from the outset, in which case there is a single wormhole solution with that charge, and relative probabilities might not matter. But string theory rather suggests that the charge might be part of the solution -- for instance, the gauge coupling may be a modulus of the higher-dimensional geometry. In such a case it makes sense to compare probabilities for different charges, i.e. for different wormholes/no-boundary instantons, see also appendix \ref{sec:Neumann} for a more detailed discussion of this point.

But then further questions arise. An immediate one is whether complex (``fuzzy'') no-boundary instantons could dominate even more \cite{Lyons:1992ua}. These, however, have a weighting that is higher for scalar field values that are lower on the potential. Hence the preference for a long inflationary phase would again be canceled by such instantons. One might than wonder whether complex wormhole equivalents also exist, which might reverse the trend for longer/shorter inflationary phases once more. A pertinent question is which of these solutions contribute to the gravitational path integral. It has recently been suggested that not all geometries and matter configurations should be summed over, as some lead to a divergence of the path integral. This is known as the Kontsevich-Segal-Witten criterion \cite{Kontsevich:2021dmb,Witten:2021nzp}, and it is trivially satisfied by our real Euclidean wormholes and by real Euclidean no-boundary instantons, but fuzzy instantons only satisfy it when the inflationary phase is sufficiently long \cite{Lehners:2022xds,Hertog:2023vot,Lehners:2023pcn}. The analogous question should then also be applied to complex wormholes, should they exist \cite{Ailiga:2026wju}. It could however also be that the asymptotic boundary conditions that form part of the framework we have analyzed here would help. If the metric is required to be real on the boundary, it may be that only real Euclidean solutions actually contribute, in which case the no-boundary solution interpolating to the top of the potential would come out as preferred.

This discussion is intimately related to the question of how to properly calculate probabilities in quantum cosmology. It is a question that remains far from resolved, as discussions in the recent literature have shown \cite{Abdalla:2026mxn,Cotler:2025gui}. Let us just remark here that the issue is not just the proper normalization of the wave function, but also which inner product is the appropriate one, and even what the correct question to ask is. Quantum mechanics (which requires us to focus on what can be measured) suggests a top-down approach in which we condition on our current knowledge and ask what came before \cite{Hawking:2002af}, rather than positing a starting point and asking how the universe likely evolved afterwards. Thus it seems that one should really calculate probabilities for past events, which is a topic that has not been explored sufficiently in quantum gravity yet. 

In the recent literature there have also been many discussions regarding the Hilbert space in quantum cosmology, see e.g. \cite{Chandrasekaran:2022cip,Chakraborty:2023yed}. We merely point out here that the framework considered in this paper, in which a de Sitter universe arises from a larger AdS background space, allows one to embed a cosmological solution (perhaps requiring a small, finite-sized Hilbert space for its description) into a larger setting. It seems worthwhile to explore this issue further, especially in the zero-charge limit in which we obtain a disconnected no-boundary instanton disconnected from the AdS background. More generally, one may ask what the consequences of the (at least naively inaccessible) background space are from the point of view of the freshly nucleated universe.

There is also the general question of what the impact of wormhole solutions is on everyday physics. Wormholes are puzzling because they can connect distant regions of the universe non-locally, potentially affecting physical laws with seemingly random effects \cite{Coleman:1988cy}. This is particularly puzzling here since we saw that wineglass wormholes also exist with flat asymptotics (and the rim has microscopic size in realistic examples), so that they may connect any two regions of the space surrounding us. Moreover, some of these solutions have an enhanced, positive weighting (i.e. a negative Euclidean action). What are we to make of this?

Could it be that these wormholes are unstable \cite{Jonas:2023qle}? But if they have negative modes, this may simply point to the existence of a more stable solution at even lower values of the Euclidean action, which would exacerbate the problem (unless the lower action solutions always happen to be disconnected ones). Or might these wormholes simply not exist? We should keep in mind that they require somewhat special matter to support them. It could be that axions do not exist, nor spherically symmetric gauge field configurations. In this context, it would be interesting to find out if less-symmetric wormholes, supported by ordinary electromagnetism, can be found. Analogously, one might want to look for wormhole solutions in related contexts, such as modified gravity theories \cite{Battarra:2014naa} or
extended Einstein-Skyrme theory \cite{Canfora:2025roy}, to investigate the robustness of our results. These are certainly worthwhile topics for future research. 

This issue is also connected to the unsolved question of integration contours for gravitational path integrals. The integration contours determine which saddles do, and which do not, contribute to the integral. For example, if the gravitational path integral is defined as a sum over purely Lorentzian metrics, then only saddles with zero or negative weighting can possibly contribute \cite{Feldbrugge:2017kzv}. But that alone would not be a sufficient explanation here, as we saw that wineglass wormholes exist even with zero or only slightly negative weighting. This brings us immediately back to a discussion of changing constants of nature, alpha vacua, etc. Only that these issues are even sharper here, given that we are in the presence of a continuous family of wormhole solutions. There remains much to be understood.

\section*{Acknowledgements}
The authors thank the Albert-Einstein-Institute for kind hospitality while this work was conducted. The work of G.L. is supported in part by the Shota Rustaveli National Science Foundation of Georgia with Grant N FR-25-1194.

\appendix

\section{Magnetic $SU(2)$ configurations} \label{sec:appendix}

It has been understood for quite some time (see \cite{Gupta:1989bs}) that magnetic charges may allow for Euclidean wormhole solutions. This is because in Euclidean time, for electric fields $E$ and magnetic fields $B,$ the energy density is of the form
\begin{align}
    \rho_E \sim E^2 - B^2\,,
\end{align}
with an extra minus sign compared to the Lorentzian setting. Thus magnetic fields contribute negatively to the Euclidean energy density, and this feature allows them to support a wormhole throat.

A complication, however, is that a standard electromagnetic $U(1)$ field necessarily breaks the symmetry of the spatial $3$-sphere configurations that we are interested in. It is then possible to use appropriate arrangements of three separate $U(1)$ fields \cite{Marolf:2021kjc} or, perhaps more fittingly, an $SU(2)$ gauge theory \cite{Hosoya:1989zn}. The idea here is to ``align'' the gauge group $SU(2)$ (which, seen as a manifold, is nothing but the $3$-sphere $S^3$) with the spherical spatial directions of the metric. In this way one obtains a stress-energy tensor that respects the isotropy of the spherical directions. Let us briefly review this construction \cite{Hosoya:1989zn}, streamlined for our purposes.

The Yang-Mills action with gauge coupling $g$ is given by
\begin{align}
   S_{YM} = \frac{1}{4g^2} \int \mathrm{d}\tau \mathrm{d}^3 x \sqrt{g} F_{\mu\nu}^a F^{\mu\nu a}\,,
\end{align}
where $a=1,2,3$ is the internal, isospin index and $F^a$ is the $SU(2)$ field strength defined in terms of the gauge potential $A^a$ via
\begin{align}
    F^a = \mathrm{d}A^a + \frac{1}{2}\epsilon^{abc} A^b \wedge A^c\,,\label{A2}
\end{align}
where the second term characterizes the non-Abelian self-interaction part of the field. We will stick with a FLRW ansatz for the metric, and we denote the spherical dreibeine via
\begin{align}
    e^i = a(\tau) \sigma^i\,,
\end{align}
where $\sigma^i$ are the left-invariant one-forms on the $S^3.$ They satisfy the Maurer-Cartan structure equations
\begin{align}
    \mathrm{d}\sigma^a + \epsilon^{abc} \sigma^b \wedge \sigma^c = 0\,.\label{A4}
\end{align}
We will write the gauge potential as
\begin{align}
    A^i = \Phi(\tau) \sigma^i\,,
\end{align}
where we have now explicitly identified frame and gauge indices. A quick calculation, using Eqs.~\eqref{A2} and \eqref{A4}, then gives
\begin{align}
    F^a = \dot\Phi \, \mathrm{d}\tau \wedge \sigma^a + \left(\frac{1}{2}\Phi^2 - \Phi \right)\epsilon^{abc}\sigma^b \wedge \sigma^c\,.
\end{align}
Thus the field strength consist of an electric component $\dot\Phi$ and a magnetic component $\left(\frac{1}{2}\Phi^2 - \Phi \right).$ To evaluate the action we require the traces (containing inverse metrics)
\begin{align}
    F^a_{\tau i}F^{\tau i a} = 3\frac{\dot\Phi^2}{a^2}\,, \qquad F^a_{ij}F^{ija} = \frac{6}{a^4}\left(\frac{1}{2}\Phi^2 - \Phi \right)^2\,,
\end{align}
leading to
\begin{align}
    S_{YM} &= \frac{1}{4g^2} \int \mathrm{d}\tau \mathrm{d}^3 x \sqrt{g} F_{\mu\nu}^a F^{\mu\nu a} \\ &= \frac{2\pi^2}{4g^2}\int \mathrm{d}\tau a^3 \left[ 3\frac{\dot\Phi^2}{a^2} +\frac{6}{a^4}\left(\frac{1}{2}\Phi^2 - \Phi \right)^2 \right]\,.
\end{align}
Varying with respect to $\Phi$ yields the equation of motion
\begin{align}
    \ddot\Phi + \frac{\dot{a}}{a}\dot\Phi - \frac{4}{a^2}\left(\frac{1}{2}\Phi^2 - \Phi \right)\left(\Phi - 1 \right) = 0\,.
\end{align}
Of particular interest to us is the purely magnetic solution $\Phi=1,$ which leads to the effective action
\begin{align}
    S_{YM} = \frac{3\pi^2}{4g^2}\int \mathrm{d}\tau \frac{1}{a}\,.
\end{align}
Upon rescaling $g$ this provides the magnetic contribution in \eqref{action}.

\section{Boundary condition for the charge} \label{sec:Neumann}

The setting that we have in mind in this work is the AdS/CFT correspondence. As such, even though we work in four dimensions, we implicitly assume that our theory under consideration in fact originates via a compactification from a higher-dimensional starting point. Therefore the axion charge (or the magnetic charge) should be regarded as being part of the higher-dimensional solution, which also implies that it is an adjustable parameter. To implement this notion in four dimensions, it is most natural to use a Neumann boundary condition for the axion (or magnetic field). 

We will illustrate this procedure explicitly with the example of the axion, described as a 2-form field $B_{\mu\nu}$ with field strength $H_{\mu\nu\rho}$ and action (see e.g. \cite{Jonas:2023ipa}) 
\begin{align}
    S_a = \frac{1}{12}H_{\mu\nu\rho}H^{\mu\nu\rho}\,,
\end{align}
where the decay constant is set to unity, $f=1,$ for simplicity here. In order to obtain a Neumann boundary condition, we add the surface term 
\begin{align}
    S_{Neumann} = -\frac{1}{2}\int \mathrm{d}^3x \sqrt{h} n_\mu B_{\nu\rho} H^{\mu\nu\rho}\,,
\end{align}
where $n_\mu$ is the normal to the boundary. The total variation of the action then includes the surface term
\begin{align}
    \delta S = \mathrm{previous \, terms}  -\frac{1}{2}\int \mathrm{d}^3x \sqrt{h}  B_{\nu\rho} \delta( n_\mu H^{\mu\nu\rho})\,.
\end{align}
Here $n_\mu = (1,0,0,0).$ Meanwhile, the $S^3$ symmetry implies that $H_{\tau ij} = 0,$ where $\tau$ is the Euclidean time and $i,j$ are coordinates on the sphere. Thus $n_\mu H^{\mu\nu\rho}=H^{\tau \nu\rho} = 0$ and we can see that the variational principle is solved due to the symmetry of the ansatz. Moreover, by the same token the boundary action vanishes. 

By contrast, the constancy of the charge arises due to the Bianchi identity, again combined with the $S^3$ symmetry ($H_{ijk} = q(\tau)\epsilon_{ijk}, \, H_{\tau ij=0}$), which reduces to
\begin{align}
    \mathrm{d} H \propto \partial_{[\mu} H_{\nu\rho\sigma ]} = 0 \qquad \to \qquad \partial_\tau q(\tau) \epsilon_{ijk} = 0\,. 
\end{align}

We note that the argument just presented regarding the Neumann surface term is rather general. If we consider a field $q$ with its canonically conjugate momentum $\pi_q,$ then the Neumann surface term (which takes the form of a Legendre transform) has the structure $\int q \pi_q.$ By construction, when varying the action one will obtain a surface term of the form $q\delta(\pi_q)=0.$ And for a constant field $q$ with a vanishing momentum $\pi_q \propto \dot{q}$ such a term will vanish, as will the value of the on-shell action.  

These arguments imply that there is no further contribution to the on-shell action coming from the Neumann boundary condition on the charge. Moreover, we note that different values of the charge all satisfy the same boundary condition, and thus it makes sense to compare probabilities for solutions with different charges. 

\bibliographystyle{utphys}
\bibliography{biblio}

@article{Coleman:1980aw,
    author = "Coleman, Sidney R. and De Luccia, Frank",
    title = "{Gravitational Effects on and of Vacuum Decay}",
    reportNumber = "SLAC-PUB-2463",
    doi = "10.1103/PhysRevD.21.3305",
    journal = "Phys. Rev. D",
    volume = "21",
    pages = "3305",
    year = "1980"
}

@article{Guth:1981ff,
    author = "Guth, Alan H.",
    title = "{The Inflationary Universe: A Possible Solution to the Horizon and Flatness Problems}",
    journal = "Phys. Rev. D",
    volume = "23",
    pages = "347--356",
    year = "1981",
    doi = "10.1103/PhysRevD.23.347"
}

@article{Linde:1981mu,
    author = "Linde, Andrei D.",
    title = "{A New Inflationary Universe Scenario: A Possible Solution of the Horizon, Flatness, Homogeneity, Isotropy and Primordial Monopole Problems}",
    journal = "Phys. Lett. B",
    volume = "108",
    pages = "389--393",
    year = "1982",
    doi = "10.1016/0370-2693(82)91219-9"
}

@article{Albrecht:1982wi,
    author = "Albrecht, Andreas and Steinhardt, Paul J.",
    title = "{Cosmology for Grand Unified Theories with Radiatively Induced Symmetry Breaking}",
    journal = "Phys. Rev. Lett.",
    volume = "48",
    pages = "1220--1223",
    year = "1982",
    doi = "10.1103/PhysRevLett.48.1220"
}

@article{Hartle:1983ai,
    author = "Hartle, J. B. and Hawking, S. W.",
    editor = "Fang, Li-Zhi and Ruffini, R.",
    title = "{Wave Function of the Universe}",
    reportNumber = "PRINT-83-0937 (CAMBRIDGE)",
    doi = "10.1103/PhysRevD.28.2960",
    journal = "Phys. Rev. D",
    volume = "28",
    pages = "2960--2975",
    year = "1983"
}

@article{Farhi:1986ty,
    author = "Farhi, Edward and Guth, Alan H.",
    title = "{An Obstacle to Creating a Universe in the Laboratory}",
    reportNumber = "MIT-CTP-1400",
    doi = "10.1016/0370-2693(87)90429-1",
    journal = "Phys. Lett. B",
    volume = "183",
    pages = "149--155",
    year = "1987"
}

@article{Lavrelashvili:1988un,
    author = "Lavrelashvili, G. and Rubakov, V. A. and Tinyakov, P. G.",
    title = "{Loss of Quantum Coherence Due to Topological Changes: A Toy Model}",
    doi = "10.1142/S0217732388001483",
    journal = "Mod. Phys. Lett. A",
    volume = "3",
    pages = "1231--1242",
    year = "1988"
}

@article{Giddings:1987cg,
title = {Axion-induced topology change in quantum gravity and string theory},
journal = {Nuclear Physics B},
volume = {306},
number = {4},
pages = {890-907},
year = {1988},
issn = {0550-3213},
doi = {https://doi.org/10.1016/0550-3213(88)90446-4},
author = {Steven B. Giddings and Andrew Strominger}
}

@article{Coleman:1988cy,
title = {Black holes as red herrings: Topological fluctuations and the loss of quantum coherence},
journal = {Nuclear Physics B},
volume = {307},
number = {4},
pages = {867-882},
year = {1988},
issn = {0550-3213},
doi = {https://doi.org/10.1016/0550-3213(88)90110-1},
author = {Sidney Coleman}
}

@article{Rubakov:1988wx,
    author = "Rubakov, V. A. and Tinyakov, P. G.",
    title = "{Gravitational Instantons and Creation of Expanding Universes}",
    doi = "10.1016/0370-2693(88)91373-1",
    journal = "Phys. Lett. B",
    volume = "214",
    pages = "334--338",
    year = "1988"
}

@article{Farhi:1989yr,
    author = "Farhi, Edward and Guth, Alan H. and Guven, Jemal",
    title = "{Is It Possible to Create a Universe in the Laboratory by Quantum Tunneling?}",
    reportNumber = "MIT-CTP-1690",
    doi = "10.1016/0550-3213(90)90357-J",
    journal = "Nucl. Phys. B",
    volume = "339",
    pages = "417--490",
    year = "1990"
}

@article{Shvedov:1996hb,
    author = "Shvedov, O. Yu.",
    title = "{On the exponentially large probability of transition through the Lavrelashvili-Rubakov-Tinyakov wormhole}",
    eprint = "gr-qc/9602049",
    archivePrefix = "arXiv",
    doi = "10.1016/0370-2693(96)00563-1",
    journal = "Phys. Lett. B",
    volume = "381",
    pages = "45--48",
    year = "1996"
}

@article{Maldacena:1997re,
    author = "Maldacena, Juan Martin",
    title = "{The Large $N$ limit of superconformal field theories and supergravity}",
    eprint = "hep-th/9711200",
    archivePrefix = "arXiv",
    reportNumber = "HUTP-97-A097, HUTP-98-A097",
    doi = "10.4310/ATMP.1998.v2.n2.a1",
    journal = "Adv. Theor. Math. Phys.",
    volume = "2",
    pages = "231--252",
    year = "1998"
}

@article{Borde:2001nh,
    author = "Borde, Arvind and Guth, Alan H. and Vilenkin, Alexander",
    title = "{Inflationary spacetimes are not past-complete}",
    journal = "Phys. Rev. Lett.",
    volume = "90",
    pages = "151301",
    year = "2003",
    doi = "10.1103/PhysRevLett.90.151301",
    eprint = "gr-qc/0110012",
    archivePrefix = "arXiv",
    primaryClass = "gr-qc"
}

@article{Marolf:2021kjc,
    author = "Marolf, Donald and Santos, Jorge E.",
    title = "{AdS Euclidean wormholes}",
    eprint = "2101.08875",
    archivePrefix = "arXiv",
    primaryClass = "hep-th",
    doi = "10.1088/1361-6382/ac2cb7",
    journal = "Class. Quant. Grav.",
    volume = "38",
    number = "22",
    pages = "224002",
    year = "2021"
}

@article{Feldbrugge:2017kzv,
    author = "Feldbrugge, Job and Lehners, Jean-Luc and Turok, Neil",
    title = "{Lorentzian Quantum Cosmology}",
    eprint = "1703.02076",
    archivePrefix = "arXiv",
    primaryClass = "hep-th",
    doi = "10.1103/PhysRevD.95.103508",
    journal = "Phys. Rev. D",
    volume = "95",
    number = "10",
    pages = "103508",
    year = "2017"
}

@article{Lehners:2023yrj,
    author = "Lehners, Jean-Luc",
    title = "{Review of the no-boundary wave function}",
    eprint = "2303.08802",
    archivePrefix = "arXiv",
    primaryClass = "hep-th",
    doi = "10.1016/j.physrep.2023.06.002",
    journal = "Phys. Rept.",
    volume = "1022",
    pages = "1--82",
    year = "2023"
}

@article{Jonas:2023ipa,
    author = "Jonas, Caroline and Lavrelashvili, George and Lehners, Jean-Luc",
    title = "{Zoo of axionic wormholes}",
    doi = "10.1103/PhysRevD.108.066012",
    journal = "Phys. Rev. D",
    volume = "108",
    number = "6",
    pages = "066012",
    year = "2023"
}

@article{Jonas:2023qle,
    author = "Jonas, Caroline and Lavrelashvili, George and Lehners, Jean-Luc",
    title = "{Stability of axion-dilaton wormholes}",
    eprint = "2312.08971",
    archivePrefix = "arXiv",
    primaryClass = "hep-th",
    doi = "10.1103/PhysRevD.109.086022",
    journal = "Phys. Rev. D",
    volume = "109",
    number = "8",
    pages = "086022",
    year = "2024"
}

@article{Betzios:2024oli,
    author = "Betzios, Panos and Papadoulaki, Olga",
    title = "{Inflationary Cosmology from Anti-de Sitter Wormholes}",
    eprint = "2403.17046",
    archivePrefix = "arXiv",
    primaryClass = "hep-th",
    doi = "10.1103/PhysRevLett.133.021501",
    journal = "Phys. Rev. Lett.",
    volume = "133",
    number = "2",
    pages = "021501",
    year = "2024"
}

@article{Betzios:2024zhf,
    author = "Betzios, Panos and Gialamas, Ioannis D. and Papadoulaki, Olga",
    title = "{Magnetic anti{\textendash}de Sitter wormholes as seeds for Higgs inflation}",
    eprint = "2412.03639",
    archivePrefix = "arXiv",
    primaryClass = "hep-th",
    doi = "10.1103/9w85-fyhs",
    journal = "Phys. Rev. D",
    volume = "111",
    number = "12",
    pages = "123542",
    year = "2025"
}

@article{Lan:2024gnv,
    author = "Lan, Qing-Yu and Piao, Yun-Song",
    title = "{Prepare inflationary universe via the Euclidean charged wormhole}",
    eprint = "2411.13844",
    archivePrefix = "arXiv",
    primaryClass = "gr-qc",
    month = "11",
    year = "2024"
}

@article{Betzios:2026rbv,
    author = "Betzios, Panos and Ghiringhelli, Paul and Gialamas, Ioannis D. and Papadoulaki, Olga",
    title = "{A Menagerie of Wormholes and Cosmologies in the Gravitational Path Integral}",
    eprint = "2602.23432",
    archivePrefix = "arXiv",
    primaryClass = "hep-th",
    month = "2",
    year = "2026"
}

@article{Balasubramanian:1998sn,
    author = "Balasubramanian, Vijay and Kraus, Per and Lawrence, Albion E.",
    title = "{Bulk versus boundary dynamics in anti-de Sitter space-time}",
    eprint = "hep-th/9805171",
    archivePrefix = "arXiv",
    reportNumber = "HUTP-98-A028, CALT-68-2171",
    doi = "10.1103/PhysRevD.59.046003",
    journal = "Phys. Rev. D",
    volume = "59",
    pages = "046003",
    year = "1999"
}

@article{deBoer:1999tgo,
    author = "de Boer, Jan and Verlinde, Erik P. and Verlinde, Herman L.",
    title = "{On the holographic renormalization group}",
    eprint = "hep-th/9912012",
    archivePrefix = "arXiv",
    reportNumber = "PUPT-1898, ITFA-99-39, SPIN-1999-29",
    doi = "10.1088/1126-6708/2000/08/003",
    journal = "JHEP",
    volume = "08",
    pages = "003",
    year = "2000"
}

@article{Elvang:2016tzz,
    author = "Elvang, Henriette and Hadjiantonis, Marios",
    title = "{A Practical Approach to the Hamilton-Jacobi Formulation of Holographic Renormalization}",
    eprint = "1603.04485",
    archivePrefix = "arXiv",
    primaryClass = "hep-th",
    doi = "10.1007/JHEP06(2016)046",
    journal = "JHEP",
    volume = "06",
    pages = "046",
    year = "2016"
}

@article{Lyons:1992ua,
    author = "Lyons, G. W.",
    title = "{Complex solutions for the scalar field model of the universe}",
    doi = "10.1103/PhysRevD.46.1546",
    journal = "Phys. Rev. D",
    volume = "46",
    pages = "1546--1550",
    year = "1992"
}

@article{Witten:2021nzp,
    author = "Witten, Edward",
    title = "{A Note On Complex Spacetime Metrics}",
    eprint = "2111.06514",
    archivePrefix = "arXiv",
    primaryClass = "hep-th",
    month = "11",
    year = "2021"
}

@article{Hertog:2023vot,
    author = "Hertog, Thomas and Janssen, Oliver and Karlsson, Joel",
    title = "{Kontsevich-Segal Criterion in the No-Boundary State Constrains Inflation}",
    eprint = "2305.15440",
    archivePrefix = "arXiv",
    primaryClass = "hep-th",
    doi = "10.1103/PhysRevLett.131.191501",
    journal = "Phys. Rev. Lett.",
    volume = "131",
    number = "19",
    pages = "191501",
    year = "2023"
}

@article{Lehners:2022xds,
    author = "Lehners, Jean-Luc",
    title = "{Allowable complex scalars from Kaluza-Klein compactifications and metric rescalings}",
    eprint = "2209.14669",
    archivePrefix = "arXiv",
    primaryClass = "hep-th",
    doi = "10.1103/PhysRevD.107.046004",
    journal = "Phys. Rev. D",
    volume = "107",
    number = "4",
    pages = "046004",
    year = "2023"
}

@article{Lehners:2023pcn,
    author = "Lehners, Jean-Luc and Quintin, Jerome",
    title = "{A small Universe}",
    eprint = "2309.03272",
    archivePrefix = "arXiv",
    primaryClass = "hep-th",
    doi = "10.1016/j.physletb.2024.138488",
    journal = "Phys. Lett. B",
    volume = "850",
    pages = "138488",
    year = "2024"
}

@article{Lavrelashvili:2026zsw,
    author = "Lavrelashvili, George and Lehners, Jean-Luc",
    title = "{Nucleating an Inflationary Universe: Euclidean Wormholes and their No-Boundary Limit}",
    eprint = "2603.11003",
    archivePrefix = "arXiv",
    primaryClass = "hep-th",
    month = "3",
    year = "2026"
}

@article{Ailiga:2026wju,
    author = "Ailiga, Manishankar and Narain, Gaurav",
    title = "{Note on KSW-allowability of Wine-Glass Geometry}",
    eprint = "2603.23457",
    archivePrefix = "arXiv",
    primaryClass = "hep-th",
    month = "3",
    year = "2026"
}

@article{Hosoya:1989zn,
    author = "Hosoya, Akio and Ogura, Waichi",
    title = "{Wormhole Instanton Solution in the Einstein {Yang-Mills} System}",
    reportNumber = "RRK-89-7, INS-733",
    doi = "10.1016/0370-2693(89)91020-4",
    journal = "Phys. Lett. B",
    volume = "225",
    pages = "117--120",
    year = "1989"
}

@article{Gupta:1989bs,
    author = "Gupta, Arun K. and Hughes, James and Preskill, John and Wise, Mark B.",
    title = "{Magnetic Wormholes and Topological Symmetry}",
    reportNumber = "CALT-68-1557",
    doi = "10.1016/0550-3213(90)90228-6",
    journal = "Nucl. Phys. B",
    volume = "333",
    pages = "195--220",
    year = "1990"
}

@article{Hawking:1981gb,
    author = "Hawking, S. W.",
    editor = "Fang, Li-Zhi and Ruffini, R.",
    title = "{The Boundary Conditions of the Universe}",
    reportNumber = "PRINT-82-0179 (CAMBRIDGE)",
    journal = "Pontif. Acad. Sci. Scr. Varia",
    volume = "48",
    pages = "563--574",
    year = "1982"
}

@article{Rudelius:2019cfh,
    author = "Rudelius, Tom",
    title = "{Conditions for (No) Eternal Inflation}",
    eprint = "1905.05198",
    archivePrefix = "arXiv",
    primaryClass = "hep-th",
    doi = "10.1088/1475-7516/2019/08/009",
    journal = "JCAP",
    volume = "08",
    pages = "009",
    year = "2019"
}

@article{Jonas:2021xkx,
    author = "Jonas, Caroline and Lehners, Jean-Luc and Quintin, Jerome",
    title = "{Cosmological consequences of a principle of finite amplitudes}",
    eprint = "2102.05550",
    archivePrefix = "arXiv",
    primaryClass = "hep-th",
    doi = "10.1103/PhysRevD.103.103525",
    journal = "Phys. Rev. D",
    volume = "103",
    number = "10",
    pages = "103525",
    year = "2021"
}

@article{Abdalla:2026mxn,
    author = "Abdalla, Ahmed I. and Antonini, Stefano and Bousso, Raphael and Iliesiu, Luca V. and Levine, Adam and Shahbazi-Moghaddam, Arvin",
    title = "{Consistent Evaluation of the No-Boundary Proposal}",
    eprint = "2602.02682",
    archivePrefix = "arXiv",
    primaryClass = "hep-th",
    month = "2",
    year = "2026"
}

@article{Hawking:2002af,
    author = "Hawking, S. W. and Hertog, Thomas",
    title = "{Why does inflation start at the top of the hill?}",
    eprint = "hep-th/0204212",
    archivePrefix = "arXiv",
    doi = "10.1103/PhysRevD.66.123509",
    journal = "Phys. Rev. D",
    volume = "66",
    pages = "123509",
    year = "2002"
}

@article{Ijjas:2013vea,
    author = "Ijjas, Anna and Steinhardt, Paul J. and Loeb, Abraham",
    title = "{Inflationary paradigm in trouble after Planck2013}",
    eprint = "1304.2785",
    archivePrefix = "arXiv",
    primaryClass = "astro-ph.CO",
    doi = "10.1016/j.physletb.2013.05.023",
    journal = "Phys. Lett. B",
    volume = "723",
    pages = "261--266",
    year = "2013"
}

@article{Anabalon:2019equ,
    author = "Anabal{\'o}n, Andr{\'e}s and Bramberger, Sebastian F. and Lehners, Jean-Luc",
    title = "{Kerr-NUT-de Sitter as an Inhomogeneous Non-Singular Bouncing Cosmology}",
    eprint = "1904.07285",
    archivePrefix = "arXiv",
    primaryClass = "hep-th",
    doi = "10.1007/JHEP09(2019)096",
    journal = "JHEP",
    volume = "09",
    pages = "096",
    year = "2019"
}

@article{Aguirre:2006ak,
    author = "Aguirre, Anthony and Gratton, Steven and Johnson, Matthew C",
    title = "{Hurdles for recent measures in eternal inflation}",
    eprint = "hep-th/0611221",
    archivePrefix = "arXiv",
    doi = "10.1103/PhysRevD.75.123501",
    journal = "Phys. Rev. D",
    volume = "75",
    pages = "123501",
    year = "2007"
}

@article{Planck:2018jri,
    author = "Akrami, Y. and others",
    collaboration = "Planck",
    title = "{Planck 2018 results. X. Constraints on inflation}",
    eprint = "1807.06211",
    archivePrefix = "arXiv",
    primaryClass = "astro-ph.CO",
    doi = "10.1051/0004-6361/201833887",
    journal = "Astron. Astrophys.",
    volume = "641",
    pages = "A10",
    year = "2020"
}

@article{Kontsevich:2021dmb,
    author = "Kontsevich, Maxim and Segal, Graeme",
    title = "{Wick Rotation and the Positivity of Energy in Quantum Field Theory}",
    eprint = "2105.10161",
    archivePrefix = "arXiv",
    primaryClass = "hep-th",
    doi = "10.1093/qmath/haab027",
    journal = "Quart. J. Math. Oxford Ser.",
    volume = "72",
    number = "1-2",
    pages = "673--699",
    year = "2021"
}

@article{Cotler:2025gui,
    author = "Cotler, Jordan and Jensen, Kristan",
    title = "{Norm of the no-boundary state}",
    eprint = "2506.20547",
    archivePrefix = "arXiv",
    primaryClass = "hep-th",
    doi = "10.1007/JHEP03(2026)180",
    journal = "JHEP",
    volume = "03",
    pages = "180",
    year = "2026"
}

@article{Chandrasekaran:2022cip,
    author = "Chandrasekaran, Venkatesa and Longo, Roberto and Penington, Geoff and Witten, Edward",
    title = "{An algebra of observables for de Sitter space}",
    eprint = "2206.10780",
    archivePrefix = "arXiv",
    primaryClass = "hep-th",
    doi = "10.1007/JHEP02(2023)082",
    journal = "JHEP",
    volume = "02",
    pages = "082",
    year = "2023"
}

@article{Chakraborty:2023yed,
    author = "Chakraborty, Tuneer and Chakravarty, Joydeep and Godet, Victor and Paul, Priyadarshi and Raju, Suvrat",
    title = "{The Hilbert space of de Sitter quantum gravity}",
    eprint = "2303.16315",
    archivePrefix = "arXiv",
    primaryClass = "hep-th",
    doi = "10.1007/JHEP01(2024)132",
    journal = "JHEP",
    volume = "01",
    pages = "132",
    year = "2024"
}

@article{Battarra:2014naa,
    author = "Battarra, Lorenzo and Lavrelashvili, George and Lehners, Jean-Luc",
    title = "{Creation of wormholes by quantum tunnelling in modified gravity theories}",
    eprint = "1407.6026",
    archivePrefix = "arXiv",
    primaryClass = "hep-th",
    doi = "10.1103/PhysRevD.90.124015",
    journal = "Phys. Rev. D",
    volume = "90",
    number = "12",
    pages = "124015",
    year = "2014"
}

@article{Canfora:2025roy,
    author = "Canfora, Fabrizio and Corral, Crist{\'o}bal and Diez, Borja",
    title = "{Euclidean AdS wormholes and gravitational instantons in the Einstein-Skyrme theory}",
    eprint = "2501.13024",
    archivePrefix = "arXiv",
    primaryClass = "hep-th",
    doi = "10.1103/PhysRevD.111.084072",
    journal = "Phys. Rev. D",
    volume = "111",
    number = "8",
    pages = "084072",
    year = "2025"
}

\end{document}